\documentclass[floatfix,prd,epsfig,nofootinbib,superscriptaddress,onecolumn,amssymb]{revtex4}
\usepackage[letterpaper,bindingoffset=0.0in,left=1in,right=1in,top=1in,bottom=1in,footskip=.25in]{geometry}

\usepackage{units}
\usepackage{slashed}
\usepackage{slashed}
\usepackage{graphicx,color}
\usepackage{epsfig}
\usepackage{subfigure}
\usepackage{epsfig}
\usepackage{dcolumn}
\usepackage{bm}
\usepackage{color}
\usepackage{enumitem}

\def\lsim{\mathrel{\rlap{\lower4pt\hbox{\hskip1pt$\sim$}}
    \raise1pt\hbox{$<$}}}         
\def\gsim{\mathrel{\rlap{\lower4pt\hbox{\hskip1pt$\sim$}}
    \raise1pt\hbox{$>$}}}         

\begin{document}

\title{COHERENT 2018 at the Spallation Neutron Source}

\author{D.~Akimov}
\affiliation{Institute for Theoretical and Experimental Physics named by A.I. Alikhanov of National Research Centre “Kurchatov Institute”, Moscow, 117218, Russian Federation}
\affiliation{National Research Nuclear University MEPhI (Moscow Engineering Physics Institute), Moscow, 115409, Russian Federation}

\author{J.B.~Albert}
\affiliation{Department of Physics, Indiana University, Bloomington, IN 47405, USA}

\author{P.~An}
\affiliation{Department of Physics, Duke University, Durham, NC 27708, USA}
\affiliation{Triangle Universities Nuclear Laboratory, Durham, NC 27708, USA}

\author{C.~Awe}
\affiliation{Department of Physics, Duke University, Durham, NC 27708, USA}
\affiliation{Triangle Universities Nuclear Laboratory, Durham, NC 27708, USA}

\author{P.S.~Barbeau}
\affiliation{Department of Physics, Duke University, Durham, NC 27708, USA}
\affiliation{Triangle Universities Nuclear Laboratory, Durham, NC 27708, USA}

\author{B.~Becker}
\affiliation{Department of Physics and Astronomy, University of Tennessee, Knoxville, TN 37996, USA}

\author{V.~Belov}
\affiliation{Institute for Theoretical and Experimental Physics named by A.I. Alikhanov of National Research Centre “Kurchatov Institute”, Moscow, 117218, Russian Federation}
\affiliation{National Research Nuclear University MEPhI (Moscow Engineering Physics Institute), Moscow, 115409, Russian Federation}

\author{M.A.~Blackston}
\affiliation{Oak Ridge National Laboratory, Oak Ridge, TN 37831, USA}

\author{A.~Bolozdynya}
\affiliation{National Research Nuclear University MEPhI (Moscow Engineering Physics Institute), Moscow, 115409, Russian Federation}

\author{A.~Brown}
\affiliation{Department of Mathematics and Physics, North Carolina Central University, Durham, NC 27707, USA}
\affiliation{Triangle Universities Nuclear Laboratory, Durham, NC 27708, USA}

\author{A.~Burenkov}
\affiliation{Institute for Theoretical and Experimental Physics named by A.I. Alikhanov of National Research Centre “Kurchatov Institute”, Moscow, 117218, Russian Federation}
\affiliation{National Research Nuclear University MEPhI (Moscow Engineering Physics Institute), Moscow, 115409, Russian Federation}

\author{B.~Cabrera-Palmer}
\affiliation{Sandia National Laboratories, Livermore, CA 94550, USA}

\author{M.~Cervantes}
\affiliation{Department of Physics, Duke University, Durham, NC 27708, USA}

\author{J.I.~Collar}
\affiliation{Enrico Fermi Institute, Kavli Institute for Cosmological Physics and
Department of Physics, University of Chicago, Chicago, IL 60637, USA}

\author{R.J.~Cooper}
\affiliation{Lawrence Berkeley National Laboratory, Berkeley, CA 94720, USA}

\author{R.L.~Cooper}
\affiliation{Department of Physics, New Mexico State University, Las Cruces, NM 88003, USA}
\affiliation{Los Alamos National Laboratory, Los Alamos, NM, USA, 87545, USA}

\author{J.~Daughhetee}
\affiliation{Department of Physics and Astronomy, University of Tennessee, Knoxville, TN 37996, USA}

\author{D.J.~Dean}
\affiliation{Oak Ridge National Laboratory, Oak Ridge, TN 37831, USA}

\author{M.~del~Valle~Coello}
\affiliation{Department of Physics, Indiana University, Bloomington, IN 47405, USA}

\author{J.A.~Detwiler}
\affiliation{Center for Experimental Nuclear
Physics and Astrophysics, and Department of Physics, University of
Washington, Seattle, WA 98195, USA}

\author{M.~D'Onofrio}
\affiliation{Department of Physics, Indiana University, Bloomington, IN 47405, USA}

\author{Y.~Efremenko}
\affiliation{Department of Physics and Astronomy, University of Tennessee, Knoxville, TN 37996, USA}

\author{S.R.~Elliott}
\affiliation{Los Alamos National Laboratory, Los Alamos, NM, USA, 87545, USA}

\author{E.~Erkela}
\affiliation{Center for Experimental Nuclear
Physics and Astrophysics, and Department of Physics, University of
Washington, Seattle, WA 98195, USA}

\author{A. Etenko}
\affiliation{National Research Centre “Kurchatov Institute”, Moscow, 123182, Russian Federation}
\affiliation{National Research Nuclear University MEPhI (Moscow Engineering Physics Institute), Moscow, 115409, Russian Federation}

\author{L.~Fabris}
\affiliation{Oak Ridge National Laboratory, Oak Ridge, TN 37831, USA}

\author{M.~Febbraro}
\affiliation{Oak Ridge National Laboratory, Oak Ridge, TN 37831, USA}

\author{N.~Fields}
\affiliation{Enrico Fermi Institute, Kavli Institute for Cosmological Physics and
Department of Physics, University of Chicago, Chicago, IL 60637, USA}

\author{W.~Fox}
\affiliation{Department of Physics, Indiana University, Bloomington, IN 47405, USA}

\author{A.~Galindo-Uribarri}
\affiliation{Oak Ridge National Laboratory, Oak Ridge, TN 37831, USA}
\affiliation{Department of Physics and Astronomy, University of Tennessee, Knoxville, TN 37996, USA}

\author{M.P.~Green}
\affiliation{Physics Department, North Carolina State University, Raleigh, NC 27695, USA}
\affiliation{Triangle Universities Nuclear Laboratory, Durham, NC 27708, USA}
\affiliation{Oak Ridge National Laboratory, Oak Ridge, TN 37831, USA}

\author{M.R.~Heath}
\affiliation{Department of Physics, Indiana University, Bloomington, IN 47405, USA}

\author{S.~Hedges}
\affiliation{Department of Physics, Duke University, Durham, NC 27708, USA}
\affiliation{Triangle Universities Nuclear Laboratory, Durham, NC 27708, USA}

\author{E.B.~Iverson}
\affiliation{Oak Ridge National Laboratory, Oak Ridge, TN 37831, USA}

\author{M.~Kaemingk}
\affiliation{Department of Physics, New Mexico State University, Las Cruces, NM 88003, USA}

\author{L.J.~Kaufman\footnote{Also at SLAC National Accelerator Laboratory, Menlo Park,\
 CA 94205, USA}}
\affiliation{Department of Physics, Indiana University, Bloomington, IN 47405, USA}

\author{S.R.~Klein}
\affiliation{Lawrence Berkeley National Laboratory, Berkeley, CA 94720, USA}

\author{A.~Khromov}
\affiliation{National Research Nuclear University MEPhI (Moscow Engineering Physics Institute), Moscow, 115409, Russian Federation}

\author{S.~Ki}
\affiliation{Department of Physics, Duke University, Durham, NC 27708, USA}
\affiliation{Triangle Universities Nuclear Laboratory, Durham, NC 27708, USA}

\author{A.~Konovalov\footnote{Also at Moscow Institute of Physics and Technology, Institutsky Lane 9, Dolgoprudny, Moscow region, 141700, Russian Federation}}
\affiliation{Institute for Theoretical and Experimental Physics named by A.I. Alikhanov of National Research Centre “Kurchatov Institute”, Moscow, 117218, Russian Federation}
\affiliation{National Research Nuclear University MEPhI (Moscow Engineering Physics Institute), Moscow, 115409, Russian Federation}

\author{A.~Kovalenko}
\affiliation{Institute for Theoretical and Experimental Physics named by A.I. Alikhanov of National Research Centre “Kurchatov Institute”, Moscow, 117218, Russian Federation}
\affiliation{National Research Nuclear University MEPhI (Moscow Engineering Physics Institute), Moscow, 115409, Russian Federation}

\author{A.~Kumpan}
\affiliation{National Research Nuclear University MEPhI (Moscow Engineering Physics Institute), Moscow, 115409, Russian Federation}

\author{L.~Li}
\affiliation{Department of Physics, Duke University, Durham, NC 27708, USA}
\affiliation{Triangle Universities Nuclear Laboratory, Durham, NC 27708, USA}

\author{W.~Lu}
\affiliation{Oak Ridge National Laboratory, Oak Ridge, TN 37831, USA}

\author{K.~Mann}
\affiliation{Physics Department, North Carolina State University, Raleigh, NC 27695, USA}
\affiliation{Triangle Universities Nuclear Laboratory, Durham, NC 27708, USA}

\author{Y.~Melikyan}
\affiliation{National Research Nuclear University MEPhI (Moscow Engineering Physics Institute), Moscow, 115409, Russian Federation}

\author{D.M.~Markoff}
\affiliation{Department of Mathematics and Physics, North Carolina Central University, Durham, NC 27707, USA}
\affiliation{Triangle Universities Nuclear Laboratory, Durham, NC 27708, USA}

\author{H.~Moreno}
\affiliation{Department of Physics, New Mexico State University, Las Cruces, NM 88003, USA}

\author{P.E.~Mueller}
\affiliation{Oak Ridge National Laboratory, Oak Ridge, TN 37831, USA}

\author{P.~Naumov}
\affiliation{National Research Nuclear University MEPhI (Moscow Engineering Physics Institute), Moscow, 115409, Russian Federation}

\author{J.~Newby}
\affiliation{Oak Ridge National Laboratory, Oak Ridge, TN 37831, USA}

\author{D.S.~Parno}
\affiliation{Department of Physics, Carnegie Mellon University, Pittsburgh, PA 15213, USA}

\author{S.~Penttila}
\affiliation{Oak Ridge National Laboratory, Oak Ridge, TN 37831, USA}

\author{G. Perumpilly}
\affiliation{Enrico Fermi Institute, Kavli Institute for Cosmological Physics and
Department of Physics, University of Chicago, Chicago, IL 60637, USA}

\author{D.~Radford}
\affiliation{Oak Ridge National Laboratory, Oak Ridge, TN 37831, USA}

\author{R.~Rapp}
\affiliation{Department of Physics, Carnegie Mellon University, Pittsburgh, PA 15213, USA}

\author{H.~Ray}
\affiliation{Department of Physics, University of Florida, Gainesville, FL 32611, USA}

\author{J.~Raybern}
\affiliation{Department of Physics, Duke University, Durham, NC 27708, USA}
\affiliation{Oak Ridge National Laboratory, Oak Ridge, TN 37831, USA}

\author{D.~Reyna}
\affiliation{Sandia National Laboratories, Livermore, CA 94550, USA}

\author{G.C.~Rich\footnote{Formerly at Department of Physics and Astronomy, University of North Carolina at Chapel Hill, Chapel Hill, NC 27599, USA and Triangle Universities Nuclear Laboratory, Durham, NC 27708, USA}}
\affiliation{Enrico Fermi Institute, Kavli Institute for Cosmological Physics and
Department of Physics, University of Chicago, Chicago, IL 60637, USA}

\author{D.~Rimal}
\affiliation{Department of Physics, University of Florida, Gainesville, FL 32611, USA}

\author{D.~Rudik}
\affiliation{Institute for Theoretical and Experimental Physics named by A.I. Alikhanov of National Research Centre “Kurchatov Institute”, Moscow, 117218, Russian Federation}
\affiliation{National Research Nuclear University MEPhI (Moscow Engineering Physics Institute), Moscow, 115409, Russian Federation}

\author{D.J.~Salvat}
\affiliation{Center for Experimental Nuclear
Physics and Astrophysics, and Department of Physics, University of
Washington, Seattle, WA 98195, USA}

\author{K. Scholberg\footnote{Corresponding author}}\email{schol@phy.duke.edu}
\affiliation{Department of Physics, Duke University, Durham, NC 27708, USA}

\author{B.~Scholz}
\affiliation{Enrico Fermi Institute, Kavli Institute for Cosmological Physics and
Department of Physics, University of Chicago, Chicago, IL 60637, USA}

\author{G. Sinev}
\affiliation{Department of Physics, Duke University, Durham, NC 27708, USA}

\author{W.M.~Snow}
\affiliation{Department of Physics, Indiana University, Bloomington, IN 47405, USA}

\author{V.~Sosnovtsev}
\affiliation{National Research Nuclear University MEPhI (Moscow Engineering Physics Institute), Moscow, 115409, Russian Federation}

\author{A.~Shakirov}
\affiliation{National Research Nuclear University MEPhI (Moscow Engineering Physics Institute), Moscow, 115409, Russian Federation}

\author{B.~Suh}
\affiliation{Department of Physics, Indiana University, Bloomington, IN 47405, USA}

\author{R.~Tayloe}
\affiliation{Department of Physics, Indiana University, Bloomington, IN 47405, USA}

\author{R.T.~Thornton}
\affiliation{Department of Physics, Indiana University, Bloomington, IN 47405, USA}

\author{I.~Tolstukhin}
\affiliation{Department of Physics, Indiana University, Bloomington, IN 47405, USA}

\author{J.~Vanderwerp}
\affiliation{Department of Physics, Indiana University, Bloomington, IN 47405, USA}

\author{R.L.~Varner}
\affiliation{Oak Ridge National Laboratory, Oak Ridge, TN 37831, USA}

\author{C.J.~Virtue}
\affiliation{Department of Physics, Laurentian University, Sudbury, Ontario P3E 2C6, Canada}

\author{J.~Yoo}
\affiliation{Department of Physics at Korea Advanced Institute of Science and Technology (KAIST)
and Center for Axion and Precision Physics Research (CAPP) at Institute for Basic Science (IBS), Daejeon, 34141, Republic of Korea}

\author{C.-H.~Yu}
\affiliation{Oak Ridge National Laboratory, Oak Ridge, TN 37831, USA}

\author{J.~Zettlemoyer}
\affiliation{Department of Physics, Indiana University, Bloomington, IN 47405, USA}

\begin{abstract} The primary goal of the COHERENT collaboration is to measure and study coherent elastic neutrino-nucleus scattering (CEvNS) using the high-power, $\mathcal{O}$(10-MeV), pulsed source of neutrinos provided by
  the Spallation Neutron Source (SNS) at Oak Ridge National Laboratory
  (ORNL).    In spite of its large cross section, the CEvNS process is challenging to observe, due to tiny energies of the resulting nuclear
  recoils.  The
  CEvNS cross section is cleanly predicted in the standard model of electroweak interactions;  hence its measurement provides a new way to test the model.  CEvNS is
  important for supernova physics and supernova-neutrino detection, and its
  measurement will enable validation of dark-matter detector background and
  detector-response models.  Precision measurement
  of CEvNS will address questions of nuclear structure as well.
 The COHERENT collaboration reported the first detection of CEvNS~\cite{Akimov:2017ade} using a CsI[Na] detector.   At present 
the collaboration is deploying four detector technologies:
a CsI[Na] scintillating crystal,  p-type point-contact germanium detectors, single-phase liquid argon, and NaI[Tl] crystals.  All detectors are located in the neutron-quiet basement of the SNS target building at distances 20-30 m from the SNS neutrino source. The simultaneous measurement in all four COHERENT detector subsystems will test the $N^2$ dependence of the cross section and search for new physics.  In addition, COHERENT is measuring neutrino-induced neutrons from charged- and neutral-current neutrino interactions on nuclei in shielding materials, which represent a non-negligible background for CEvNS as well as being of intrinsic interest. The Collaboration is planning as well to look for charged-current interactions of relevance to supernova and weak-interaction physics.  This document describes concisely the COHERENT physics motivations, sensitivity and next plans for measurements at the SNS to be accomplished on a few-year timescale.

\end{abstract}

\maketitle

\tableofcontents

\def\etal{{\it et al.}}
\def\etalmacro{\etal\ (\macro)}

\thispagestyle{empty}

\normalsize
\section{Introduction}\label{sec:intro}

\begin{figure}[htb]
\begin{center}
\includegraphics[width=4in]{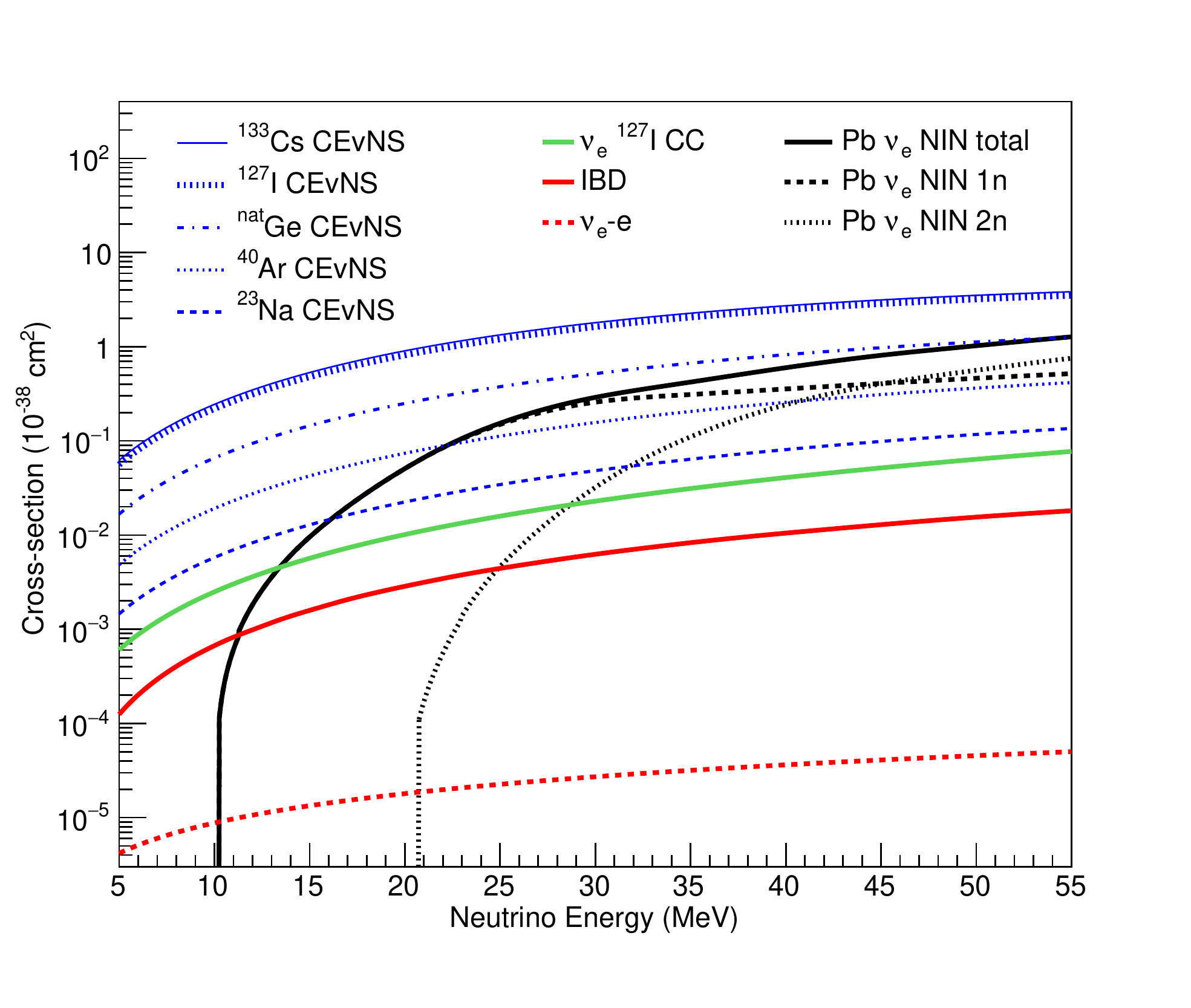}
\caption{\label{fig:cohall} 
Neutrino interaction cross sections per target as a function of neutrino energy for COHERENT target materials, as well as NIN cross sections on lead (see Sec.~\ref{sec:nins}).  Also shown, to compare with other cross sections in this energy range, are the $\nu_e$ CC cross section on $^{127}$I~\cite{Formaggio:2013kya,Distel:2002ch} and the frequently-used cross sections for inverse beta decay of $\bar{\nu}_e$ on free protons (IBD) and elastic scattering of $\nu_e$ on electrons (per electron).
}
\end{center}
\end{figure}

Coherent elastic neutrino-nucleus scattering (CEvNS\footnote{Note there exist a number of abbreviations for this process in the literature, e.g., CNS, CNNS, CENNS.  We favor a version with ``E'' for ``elastic'' to distinguish the process from inelastic coherent pion production, which is commonly confused with CEvNS by members of the high energy physics community.  We prefer to replace the first ``N'' with ``v'', for ``neutrino''  because ``NN'' means ``nucleon-nucleon'' to many in the nuclear physics community.  Finally, the Roman letter ``v'' standing in for the Greek letter ``$\nu$'' can be pronounced as a ``v''. }) was predicted in 1974 as a consequence of the neutral weak current~\cite{PhysRevD.9.1389,Kopeliovich:1974mv}.   Although the cross section is large compared to other neutrino-matter interactions in the $\sim$ 100~MeV energy range (see Fig.~\ref{fig:cohall}),
this SM process took so long to be observed due to the daunting technical requirements: very low nuclear recoil energy thresholds, intense sources/large target masses, and low backgrounds. 
Employing state-of-the-art low-energy-threshold detector technology coupled with the intense stopped-pion neutrino source available
at the Spallation Neutron Source (SNS) at Oak Ridge National Laboratory (ORNL), 
the COHERENT Collaboration has made the first measurement of 
 CEvNS~\cite{Akimov:2017ade}, using neutrinos from pion decay at rest~\cite{Drukier:1983gj, Scholberg:2005qs}.  The Collaboration has employed the data to search for physics beyond the standard model (SM).
For the completion of the current phase, a suite of four detector subsystems (CsI[Na] scintillating crystals, p-type point-contact germanium detectors, a single-phase argon detector, and an array of NaI[Tl] scintillator detectors) will be deployed in the basement of the SNS, taking advantage of decades of detector development in the dark-matter direct-detection community.  
The immediate experimental impacts of the measurements envisioned with this detector suite include:

\begin{itemize}
\item Test of the SM prediction of proportionality of the CEvNS cross section to neutron number squared, $N^2$.
\item A precision cross-section measurement on multiple targets to test for non-standard neutrino interactions, for which the interaction depends on the quark makeup of the nucleus~\cite{Scholberg:2005qs,Barranco:2007tz,Coloma:2017egw}.   First constraints have already been made with the CsI[Na] measurement~\cite{Akimov:2017ade,Coloma:2017ncl,Liao:2017uzy,Dent:2017mpr,Kosmas:2017tsq} and future possibilities are described as well in reference~\cite{Coloma:2017egw}.
\item Systematic characterization of low-threshold recoil detectors with neutrinos to validate experimental background and detector-response models, given that
CEvNS of solar and atmospheric neutrinos is an irreducible background for dark matter WIMP (Weakly Interacting Massive Particle) searches~\cite{billard:2014}.
\item Low-energy neutrino-scattering-based measurements of nuclear neutron distributions.  An interpretation of the CsI[Na] measurement is given in Ref.~\cite{Cadeddu:2017etk}.

\end{itemize}

Furthermore, the CEvNS process has one of the largest cross sections relevant for supernova dynamics and plays a significant role in core-collapse processes~\cite{Wilson:1974zz,Sato:1975id,Horowitz:2004pv}, and therefore should be measured to validate models of core-collapse supernovae.   Dark-matter detectors will also be able to detect CEvNS interactions in the event of a nearby core-collapse supernova burst~\cite{Horowitz:2003cz,Chakraborty:2013zua}, which will be sensitive to the full flavor content of the supernova signal.

As secondary goals, COHERENT will perform measurements of the charged-current (CC) and neutral-current (NC) cross sections Pb($\nu_e$,n), Fe($\nu_e$,n), and Cu($\nu_e$,n), which result in the emission of background-inducing fast neutrons.
The measurement of this cross section on lead has implications for supernova neutrino detection in the HALO supernova neutrino detector~\cite{Duba:2008zz}.
These ($\nu$, n) interactions may also influence nucleosynthesis in certain astrophysical environments~\cite{woosley1990:nuProcess,qian1997:NINSnucleosynthesis}.
 In addition, the NaI[Tl] detector array can be used to observe the CC reaction $^{127}$I($\nu_e$,e$^-$)$^{127}$Xe~\cite{Haxton:1988bk}, which has relevance both for supernova physics and for $g_A$-quenching studies (see Sec.~\ref{sec:nai}).

This paper describes the updated experimental configuration since Ref.~\cite{Akimov:2015nza}, current status, near-future plans, and briefly,
potential farther-future plans.  

	\section{Coherent Elastic Neutrino-Nucleus Scattering Physics}

The coherence of the CEvNS process results in an enhanced neutrino-nucleus cross section that is approximately proportional to $N^2$, the square of the number of neutrons in the nucleus, due to the small weak charge of the proton. The coherence condition, in which the neutrino scatters off all nucleons in a nucleus in phase with each other, requires that the wavelength of the momentum transfer be larger than the size of the target nucleus.  In this NC elastic process, initial and final states are the same.  Substantial coherence requires low-energy neutrinos (typically $<$ 50~MeV for medium-$A$ nuclei); as a result, the experimental signature for the process is a difficult-to-detect keV to sub-keV nuclear recoil, depending on the nuclear mass and neutrino energy.

\subsection{Measurement of the $N^2$-Dependent Cross Section}

\begin{figure}[htb]
\begin{center}
\includegraphics[width=5in]{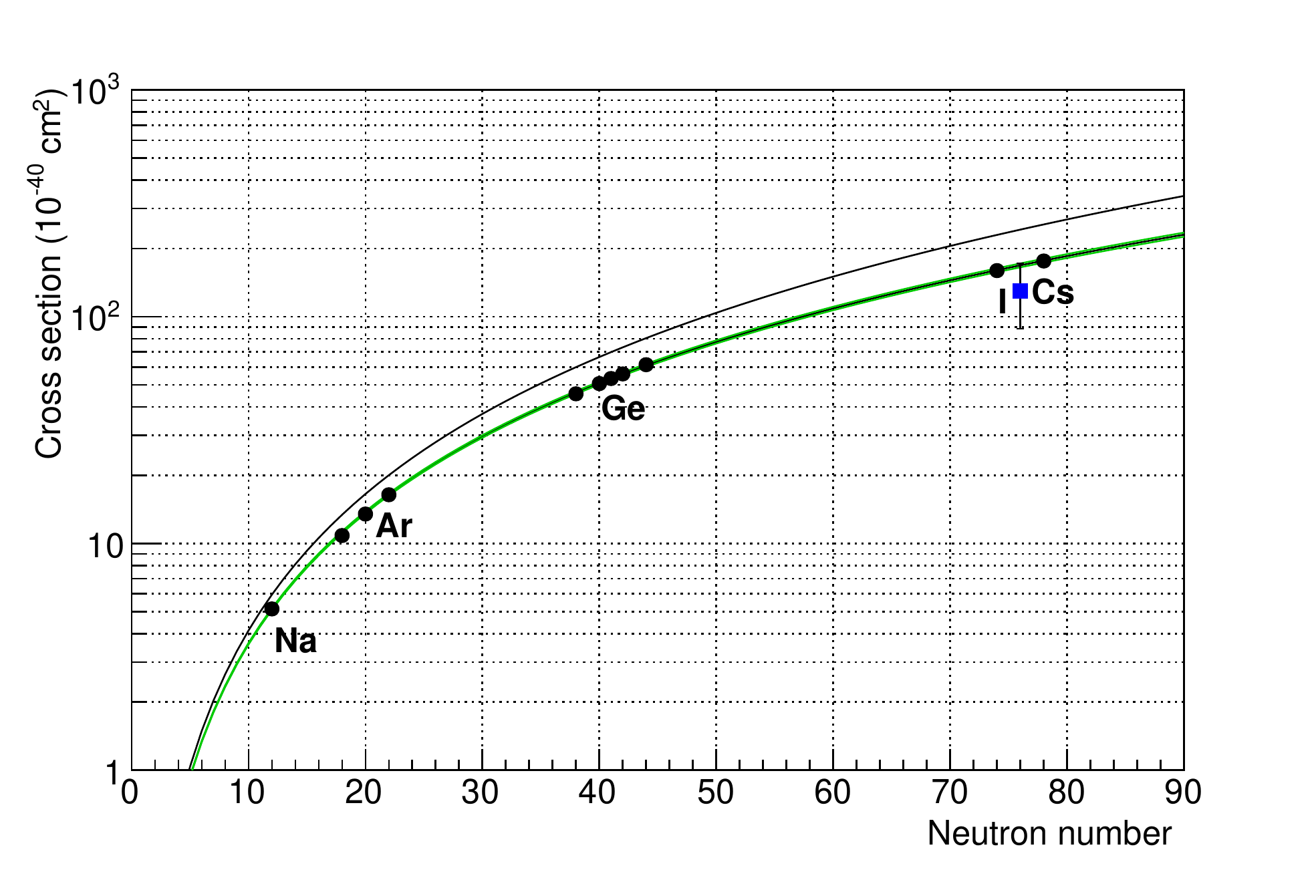}
\caption{\label{fig:cross-sections} 
Illustration of the $\propto N^2$ proportionality of the stopped-pion-neutrino flux-averaged CEvNS cross section versus neutron number $N$. The black line assumes a unity form-factor. The green band shows the effect of an assumed  form factor (very similar effects from ~\cite{Helm:1956dv, Hofstadter:1956qs} and ~\cite{Klein:1999qj}, the latter used in the prediction in Ref.~\cite{Akimov:2017ade}), with its width indicating the effect of a $\pm 3\%$ uncertainty on the assumed neutron rms radius in the Helm parameterization.  The points show the the relevant isotopes of COHERENT target materials.  The blue square shows the flux-averaged cross section inferred from the measurement reported in Ref.~\cite{Akimov:2017ade}.
}
\end{center}
\end{figure}

The cross section for CEvNS can be written as:
\begin{eqnarray}\label{eq:sevens}
\frac{d\sigma}{dT}_{coh} &=& \frac{G_F^2 M}{2\pi}\left[(G_V + G_A)^2 + (G_V - G_A)^2\left(1-\frac{T}{E_{\nu}}\right)^2 - (G_V^2 - G_A^2)\frac{MT}{E_{\nu}^2}\right] \\
G_V &=& (g_V^p Z + g_V^n N)F_{\rm nucl}^V(Q^2)\\
G_A &=& (g_A^p(Z_+ - Z_-) + g_A^n(N_+ - N_-))F_{\rm nucl}^A(Q^2),
\end{eqnarray}
where $G_F$ is the Fermi constant, $M$ is the nuclear mass, $T$ is the recoil energy, $E_{\nu}$  is the neutrino energy, 
$g_V^{n,p}$ and $g_A^{n,p}$ are vector and axial-vector coupling factors, respectively, for protons and neutrons, 
$Z$ and $N$ are the proton and neutron numbers, $Z_{\pm}$ and $N_{\pm}$ refer to the number of up or down nucleons, and $Q$ is the momentum transfer~\cite{Barranco:2005yy}.   The maximum recoil energy for a given target species and neutrino energy is $T_{\rm max} = \frac{2 E_\nu^2}{M+2 E_\nu}$.  The form factors $F_{\rm nucl}^{A,V}(Q^2)$ are point-like ($F(Q^2) = 1$) for interactions of low-energy neutrinos $<10$~MeV, but suppress the interaction rate as the wavelength of the momentum transfer becomes comparable to the size of the target nucleus (i.e.,~for higher neutrino energies and for heavier targets).   
The vector couplings appearing in $G_V$ and $G_A$ are written as:
\begin{eqnarray}
g_V^p &=& \rho_{\nu N}^{NC}\left(\frac{1}{2} - 2\hat{\kappa}_{\nu N}\sin^2 \theta_{W}\right) + 2\lambda^{uL} + 2\lambda^{uR} + \lambda^{dL} + \lambda^{dR} \\
g^n_V &=& -\frac{1}{2}\rho_{\nu N}^{NC} + \lambda^{uL}+\lambda^{uR} + 2\lambda^{dL} + 2\lambda^{dR},
\end{eqnarray}
where $\rho_{\nu N}^{NC}$, $\hat{\kappa}_{\nu N}$ are electroweak parameters,  $\lambda^{uL}, \lambda^{dL}, \lambda^{dR}, \lambda^{uR}$ are radiative corrections given in Refs.~\cite{Barranco:2005yy,PDG2014}, and $\theta_W$ is the weak mixing angle.  Figure~\ref{fig:cohall} shows CEvNS cross sections as a function of neutrino energy, and  Fig.~\ref{fig:cross-sections} shows the expected CEvNS cross section weighted by stopped-pion neutrino flux (see Section~\ref{sec:flux}), as a function of $N$, with and without form-factor suppression.
The deployment of the COHERENT detector suite in ``Neutrino Alley'', a basement location at the SNS, which is $\sim$20 m from the source of neutrinos, has resulted in one measurement~\cite{Akimov:2017ade}, and measurements with additional targets
will result in a clear observation of the coherent $N^2$ nature of the cross section (Fig.~\ref{fig:cross-sections}). 
The expected precisions of the cross section measurements will quickly become dominated by the systematic uncertainty of the knowledge of the nuclear recoil detector thresholds (see Sec.~\ref{sec:status}) and neutrino flux uncertainties.  Threshold uncertainties are dominant for the heavier Cs and I nuclei due to the lower average recoil energies for these species.

\begin{figure}[ht]
\centering
\includegraphics[height=4.0in]{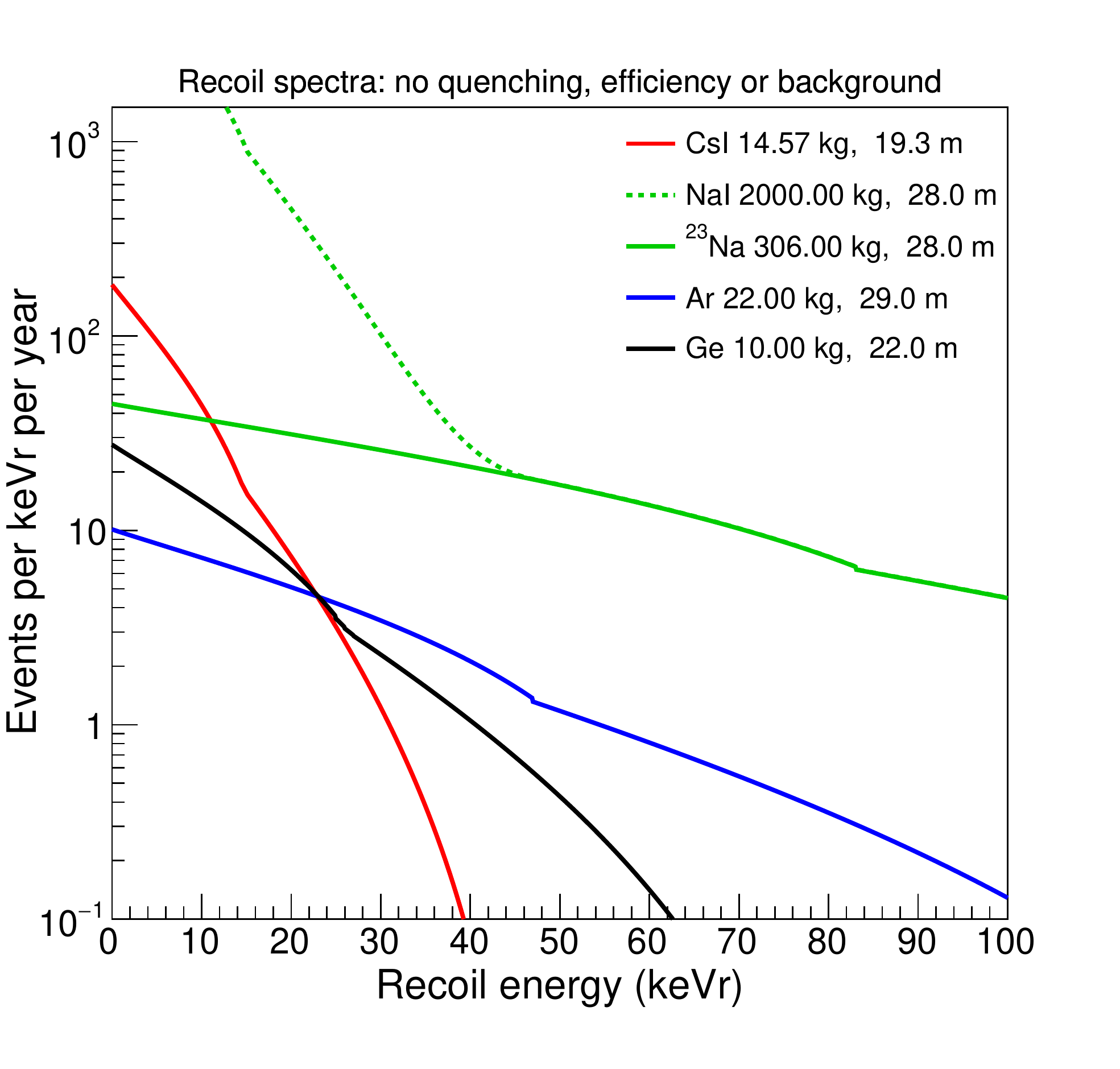}
\caption{Differential recoil rates for the COHERENT suite of detectors in Neutrino Alley.   This plot represents interaction rates for the first 6000 ns after each beam pulse, including all flavors and assuming 100\% detection efficiency.   Note this does not take into account quenching factors, which are different for different detectors, or other detector-specific efficiencies.  The kinks are due to different endpoints of prompt and delayed flavor components for different target isotope components.   The contribution from $^{23}$Na to NaI is shown separately, as the small quenching factor for $^{127}$I in NaI~\cite{Collar:2013gu}  is likely to strongly suppress the $^{127}$I contribution to the observed rate.  The materials are assumed to have natural abundances of isotopes.  
}
\label{fig:spectrum}
\end{figure}

The expected recoil spectra before detection-efficiency corrections are shown in Fig.~\ref{fig:spectrum}.

\subsection{Beyond-the-Standard-Model Physics Searches}\label{sec:bsm}

\begin{figure}[ht]
\centering
\includegraphics[width=0.45\linewidth]{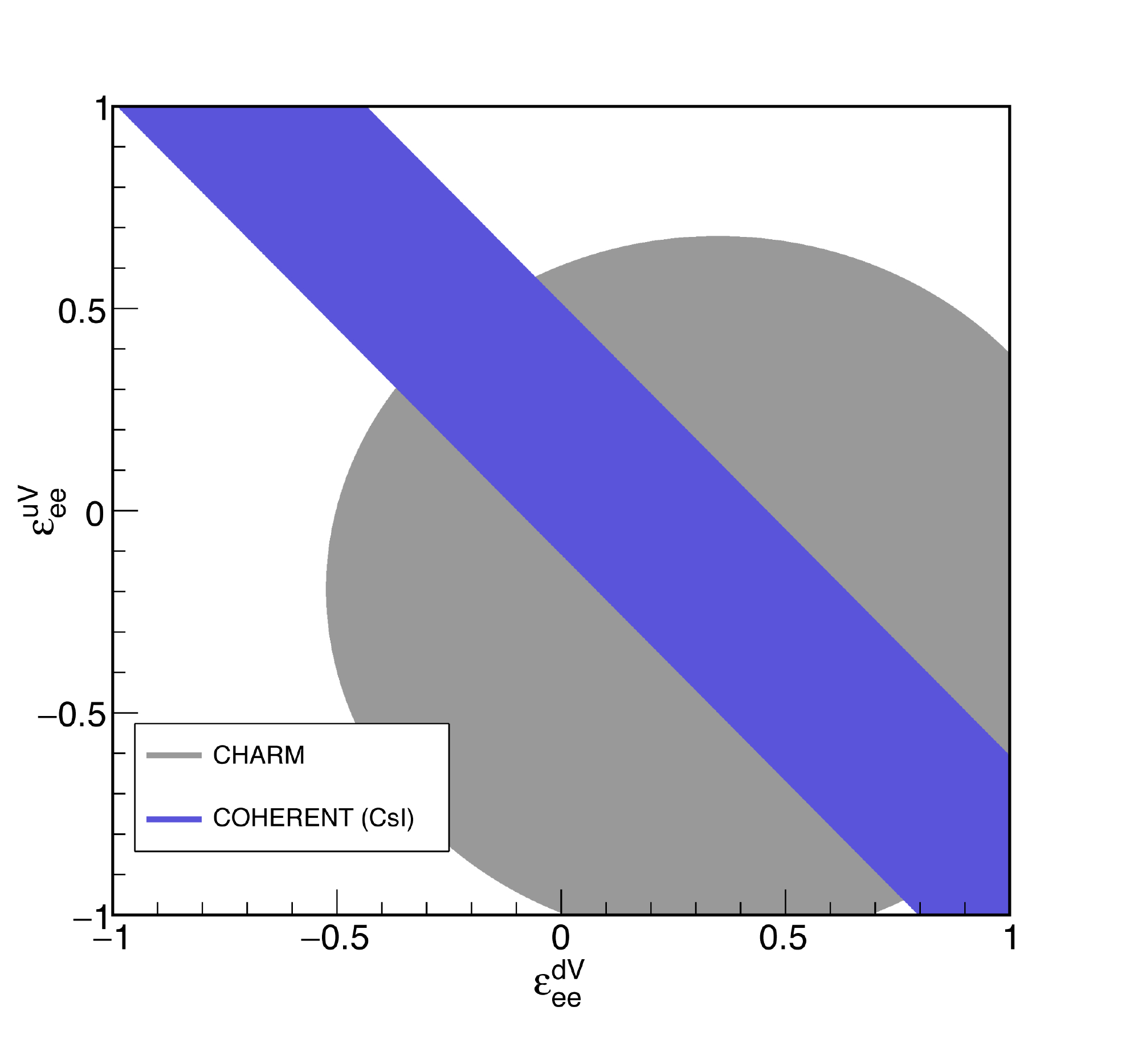}\\
\includegraphics[width=0.45\linewidth]{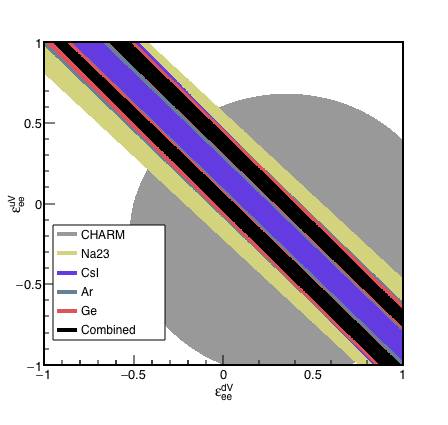}
\includegraphics[width=0.45\linewidth]{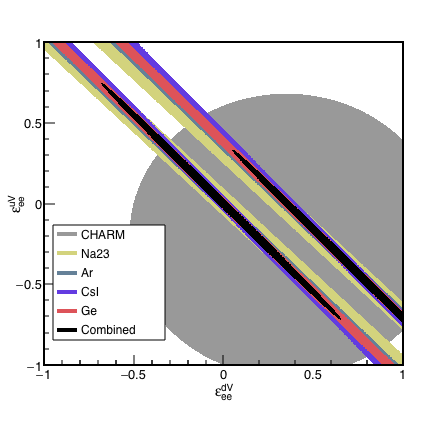}
\caption{\label{fig:NSI}Top: result from ~\cite{Akimov:2017ade} with initial constraints on two of the NSI $\epsilon$ parameters, showing also the constraint from the CHARM experiment {\protect\cite{Dorenbosch:1986tb}}.  Bottom left: same 
 with realistic assumptions for COHERENT's detector suite for the next three years.
Assumptions on mass, distance and threshold are given in Table~\ref{tab:detectors} (assuming 2~tonnes of NaI, $^{23}$Na component only, first 1000 ns of the beam window for Na and first 6000 ns for the others). 
The black shows the result from a combined fit.   Bottom right:
predicted sensitivity obtained with the COHERENT detector materials, with 5\% uncertainties on flux and event rates and assuming systematically-limited measurements. }
\end{figure}

Because the CEvNS cross section is cleanly predicted in the SM, 
deviations can indicate new physics (e.g.,~\cite{Krauss:1991ba,Barranco:2005yy, Barranco:2007tz, Harnik:2012ni,dutta2015}). As one example of a test of beyond-the-SM physics,  the CEvNS cross section for even-even nuclei, incorporating possible non-standard-interaction (NSI) neutral currents, can be parameterized as:
\begin{eqnarray}
\frac{d\sigma}{dT}_{coh} &=& \frac{G_f^2 M}{2\pi}G_V^2\left[1 + \left(1-\frac{T}{E_{\nu}}\right)^2 - \frac{MT}{E_{\nu}^2}\right]\\
G_V &=& ((g_V^p + 2\epsilon_{ee}^{uV} + \epsilon_{ee}^{dV})~Z + (g_V^n + \epsilon_{ee}^{uV} + 2\epsilon_{ee}^{dV}~)N)~F_{\rm nucl}^V
(Q^2),
\end{eqnarray}

\noindent
where the $\epsilon$'s represent new couplings~\cite{Scholberg:2005qs, Barranco:2005yy}; hence CEvNS is a novel probe of new mediators.
Neutrino-scattering constraints on the magnitude of non-zero values for $\epsilon_{ee}^{qV}$ from CHARM~\cite{Dorenbosch:1986tb} are of order unity; they are shown in Fig.~\ref{fig:NSI} as the shaded grey region\footnote{Note that these constraints are valid only for mediators not much lighter than the electroweak scale~\cite{Coloma:2017egw}.}.  A further search for NSI can be performed by comparing measured CEvNS cross sections to SM expectations.  CEvNS constraints may also help to resolve NSI ambiguities for
interpretation of neutrino oscillation parameter measurements~\cite{Coloma:2016gei, Coloma:2017egw}.
The initial NSI result from COHERENT~\cite{Akimov:2017ade} for two parameters,  $\epsilon_{ee}^{uV}$ and $\epsilon_{ee}^{dV}$ (assuming all other $\epsilon$ parameters are zero) is shown in Fig.~\ref{fig:NSI} as a blue band. Reference~\cite{Coloma:2017ncl} shows that the first data set can already constrain the ``LMA-D'' degeneracy.  Sensitivity to NSI parameters can be improved with simultaneous measurements of the cross sections on different nuclei that factor out the neutrino flux uncertainty.
The angles of the diagonal-band allowed regions vary slightly between the different isotopes due to different $N:Z$ ratios. 
With realistic parameters for the current phase of COHERENT, described in the section below,  the expected constraints from a null search for CsI[Na], Ge, Ar, NaI[Tl], and the combined analysis, are shown in Fig.~\ref{fig:NSI} as superimposed diagonal bands, along with a projection for
 long-term scenario with small systematic uncertainties to illustrate future potential.
CEvNS measurements using stopped-pion neutrinos can also constrain other NSI parameters.  Precision measurements of the CEvNS recoil spectrum also provide information on potential beyond-the-SM couplings~\cite{Liao:2017uzy}.   Examples of constraints from the CsI[Na] data set, and future potential,  are described further in Refs.~\cite{Coloma:2017ncl,Liao:2017uzy, Dent:2017mpr}.

\begin{figure}[ht]
\centering
\includegraphics[width=0.6\linewidth]{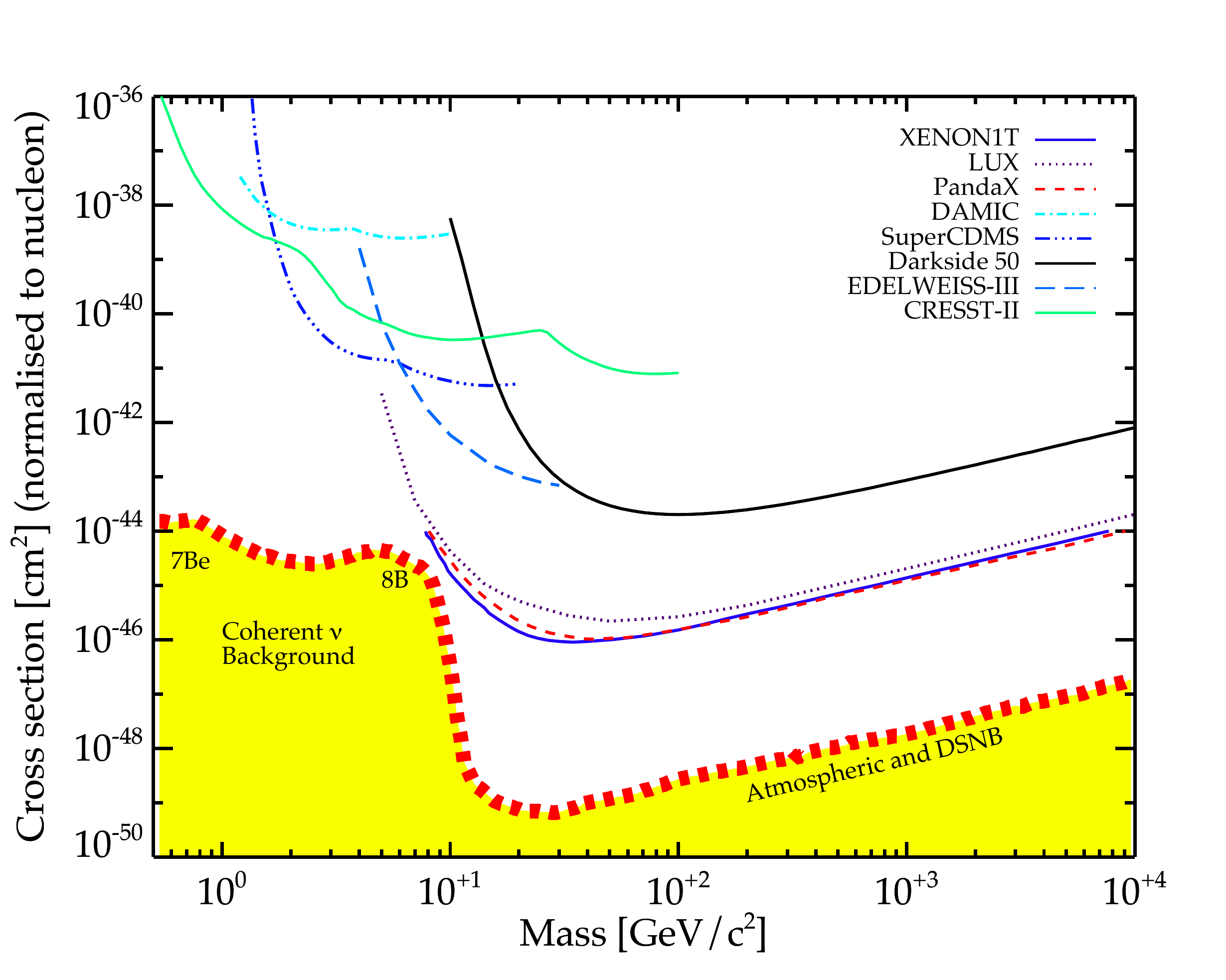}

\caption{WIMP dark-matter search parameter space, showing ``neutrino floor'' from CEvNS of solar, diffuse relic and atmospheric neutrinos as a thick dashed orange line, relevant for spin-independent WIMP-nucleon interactions. (Figure courtesy of L. Strigari, updated from~{\protect\cite{billard:2014}}.)}
\label{fig:wimps}
\end{figure}

\subsection{Relevance for Direct Dark Matter Detection Experiments}

CEvNS has long been closely linked to direct dark matter searches~\cite{Drukier:1983gj,Cabrera:1984rr}.
The CEvNS of solar and atmospheric neutrinos, which produce single-scatter recoils identical to those expected from WIMPs, is recognized as an irreducible background for dark-matter WIMP searches for next-generation dark-matter experiments~\cite{Monroe:2007xp,Gutlein:2010tq,cushman2013:snowmassDM,anderson:2011cevnsDM, billard:2014,Gutlein:2014gma}; see Fig.~\ref{fig:wimps}. Large dark-matter detectors may eventually be able to address solar neutrino physics using CEvNS~\cite{Drukier:1983gj,Billard:2014yka}.

The four detector technologies and materials proposed within the COHERENT program overlap well with those in use by the WIMP community.  Existing and next-generation experiments such as 
DarkSide~\cite{Aalseth:2017fik} and DEAP~\cite{Amaudruz:2017ekt} use argon detectors;
SuperCDMS~\cite{superCDMS:2014PRL}, EDELWEISS~\cite{Armengaud:2017rzu}, and CDEX~\cite{Zhao:2016dak} use germanium detectors; DAMA/LIBRA~\cite{Bernabei:2008yh}, DMIce~\cite{Cherwinka:2014xta}, SABRE~\cite{Tomei:2017rkg} and COSINE~\cite{Adhikari:2017esn} use NaI detectors; and the KIMS collaboration is conducting a WIMP search with CsI[Tl] crystals~\cite{kims:2012PRL}.
By utilizing the intense SNS neutrino source, COHERENT can provide detector-specific response information for CEvNS interactions as a supplement to other low-energy calibration techniques. 
A further constraint on cross-section uncertainty is afforded by the presence of higher-energy ($\gsim 10$ MeV) neutrinos in the SNS beam: these neutrinos will be a unique probe of nuclear form factors.

\subsection{Future Physics with CEvNS}\label{sec:future}

Beyond the scope of this first suite of detectors, a successful CEvNS program will enable new endeavors addressing further physics questions in the long term.  

Of particular note is that 
the weak mixing angle ($\theta_{W}$) appears in the vector coupling factors of Eq.~\ref{eq:sevens}. 
It has long been predicted that a precision measurement of the CEvNS cross section would provide a sensitive test of the weak nuclear charge and of physics above the weak scale~\cite{Krauss:1991ba}.  Theoretical uncertainties are very small.
In addition to the terms in Eq.~\ref{eq:sevens}, there also exist axial-vector terms from strange quark and weak magnetism contributions for non-even-even nuclear targets that have larger theoretical uncertainties (as much as 5\% uncertainty for light nuclei for larger $Q$). For the even-even targets we are considering, these uncertainties are $<$ 0.1\%.  For $^{23}$Na (and possibly other targets), however, the axial contribution may be measurable.
While the total neutrino flux uncertainty is estimated to be $\sim$10\%,  improvements using multiple targets to cancel uncertainties will permit few-percent or better measurements of $\sin^2\theta_W$.

CEvNS offers a measurement of the weak mixing angle at $Q\sim 40$ MeV/c,
where there is sensitivity to models of dark Z bosons, which offer a possible explanation for the $(g-2)_{\mu}$ anomaly~\cite{muonG-2:2006finalReport,jegerlehner:2009muonG-2} and a glimpse into dark sector physics~\cite{davoudiasl2012:muonAnomaly,davoudiasl2012:darkZimplications,davoudiasl2014:darkBosons,kumar2013:lowEWMA}. 
A CEvNS measurement interpreted simply as a determination of $\sin^2 \theta_W$ has substantially larger uncertainty than those from parity-violating electron scattering experiments (see~\cite{erler2014:eScatReview} for a comprehensive review).  However, CEvNS measurements will be subject to different systematic uncertainties and provides an independent reckoning at comparable values of momentum transfer.  Furthermore,  the $Q^2$ dependence of the CEvNS recoil spectrum offers 
a stringent test of models explaining the $(g-2)_{\mu}$ anomaly, with the first CsI[Na] measurement already ruling out an extensive range of currently-allowed parameters~\cite{Liao:2017uzy}.

Additional examples of future CEvNS physics include:
\begin{itemize}
\item A measurement of the CEvNS interaction spectrum is sensitive to the magnitude of the neutrino magnetic moment~\cite{Dodd:1991ni,Scholberg:2005qs,Kosmas:2015sqa} resulting in a characteristic enhancement of low-energy recoils. Because the neutrino magnetic moment is predicted to be extremely tiny in the SM, a measurement of non-zero magnetic moment would indicate new physics.  A stopped-pion flux allows access to measurements with $\nu_\mu$ flavor~\cite{Scholberg:2005qs}.

\item Precision measurement of the ratio of the muon and electron neutrino cross sections ($\sim$few~\%), which can be done by comparing prompt ($\nu_\mu$) and delayed $\nu_e$ stopped-pion signals, will be sensitive to the effective neutrino charge radius~\cite{Papavassiliou:2005cs}.  
\item Development of this flavor-blind NC CEvNS detection capability will provide a natural tool to search for oscillations into sterile neutrinos in near and far detectors~\cite{Drukier:1983gj,Giunti:2006bj}.

\item The neutron distribution function (nuclear form factor) can be probed with measurements of the CEvNS recoil spectrum~\cite{Amanik:2009zz,Patton:2012jr}. 
Incoherent contributions to the cross section are also of interest for study of nuclear axial structure~\cite{Moreno:2015bta}.   Reference~\cite{Cadeddu:2017etk} determines the neutron rms radius for CsI to 18\% using the 2017 CsI[Na] result, although without taking into account spectral shape uncertainties.

\item A $\mathcal{O}$(1 tonne) Ar or other future COHERENT recoil-sensitive detector would be sensitive to sub-GeV particles postulated in some models as candidate dark matter~\cite{deNiverville:2015mwa, Ge:2017mcq}.  Such light, weakly-coupled particles could be produced via pion decays in the 1-MW proton target and would interact coherently with nuclei in the detector. 

\end{itemize}

\section{The COHERENT Experiment}\label{sec:methods}

The COHERENT Collaboration has assembled a suite of detector
technologies suitable for the observation of CEvNS at the SNS. The SNS is
the highest-flux pulsed, stopped-pion neutrino source currently
available. This effort leverages the technological development within
the dark-matter direct-detection community over the last decade by
deploying four mature low-threshold/low-background technologies (with
five elemental targets) capable of
observing low-energy nuclear recoils: CsI[Na] scintillating crystal, p-type
point-contact (PPC) germanium detectors, single-phase liquid argon,
and NaI[Tl] scintillating crystals.
The use of targets with widely varying numbers of neutrons
provides a ready test of the $N^2$ nature of the cross section and 
allows a convenient cross-check on the measured cross
section given that detectors are subject to systematic 
response (quenching factor, QF) uncertainties.
All four detector subsystems inhabit the Neutrino Alley
hallway where a background measurement campaign (see
Sec.~\ref{sec:backgroundstudies}) has indicated very low beam-related
neutron backgrounds.  In this section we will describe the properties
of the SNS neutrino source, the results of background measurements, and the detectors.

\subsection{Neutrinos at the Spallation Neutron Source}\label{sec:flux}

A stopped-pion beam has several advantages for CEvNS detection. First, the relatively high energies enhance the cross section ($\propto E^2$) while still benefiting from coherence;  
cross sections at stopped-pion energies (up to 50~MeV) are about two orders of magnitude higher than at reactor energies ($\sim $3 MeV).   Second,
recoil energies (few to tens of keV) bring
detection of CEvNS within easy reach of the current generation of low-threshold
detectors. 
Finally, the different flavor content of the SNS flux ($\nu_\mu$, $\nu_e$ and $\bar{\nu}_\mu$) means 
that 
physics sensitivity is complementary to that for reactor experiments (which have $\bar{\nu}_e$ only). 

\begin{figure}[ht]
\centering
\subfigure[SNS neutrino energy spectrum.]{%
\includegraphics[width=0.45\linewidth]{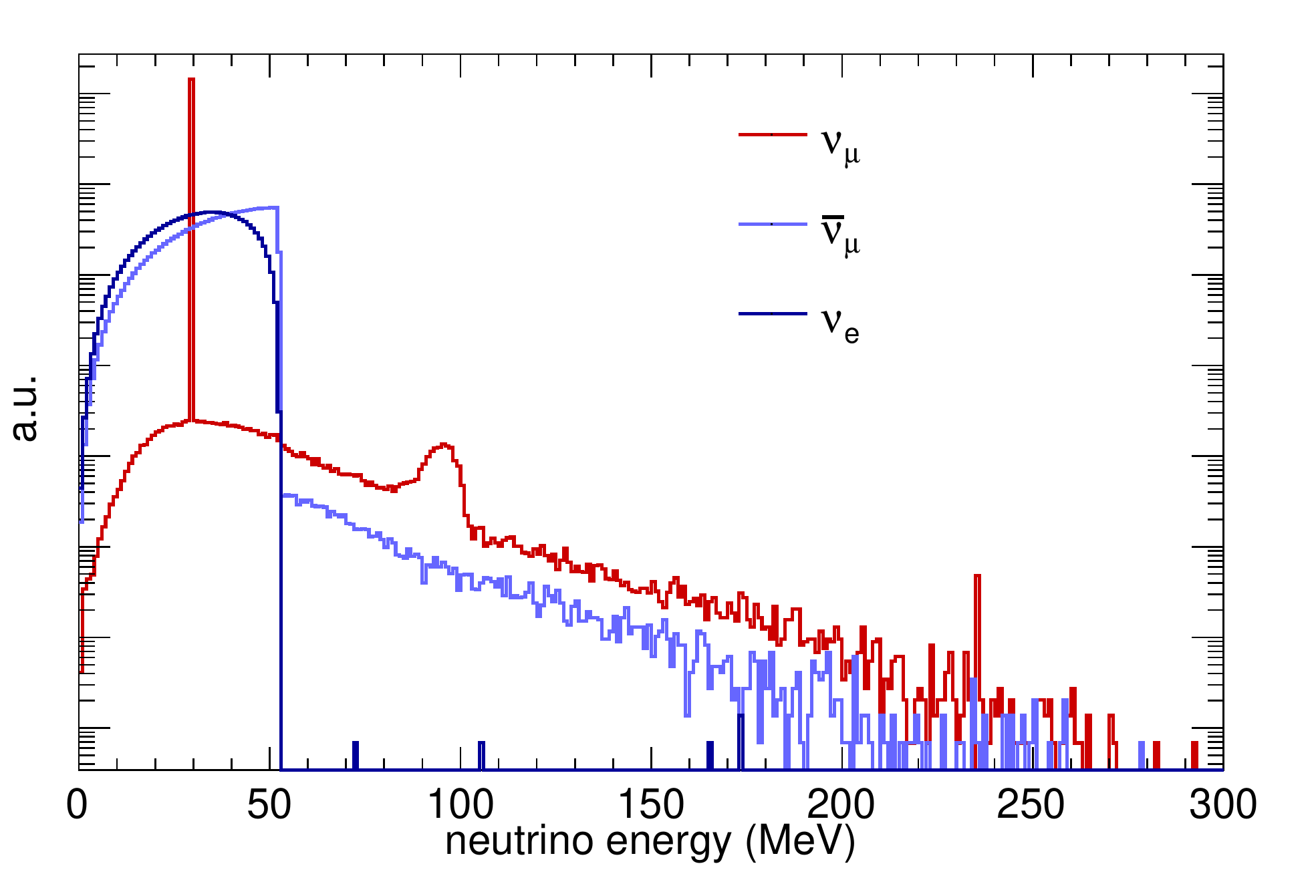}
\label{fig:snsisawesomea}}
\quad
\subfigure[SNS neutrino timing distribution.]{%
\includegraphics[width=0.45\linewidth]{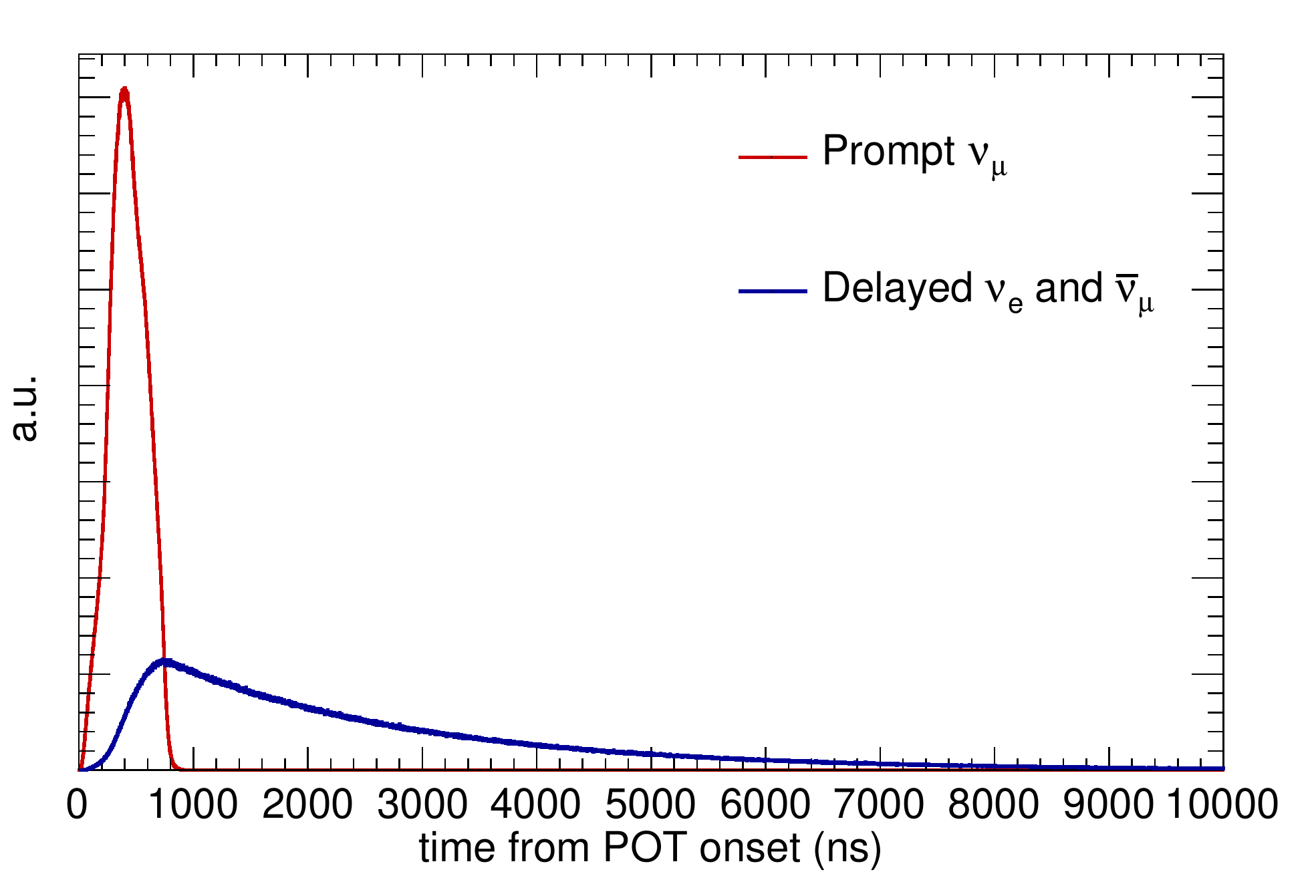}
\label{fig:snsisawesomeb}}
\caption{(a) Expected $\nu$ spectrum at the SNS, showing the very low level of decay-in-flight and other non-decay-at-rest flux, in arbitrary units; the integral is $4.3 \times 10^7$ neutrinos/cm$^2$/s at 20~m. (b) Time structure for prompt  and delayed neutrinos due to the 60-Hz pulses. ``Prompt'' refers to neutrinos from pion decay, and ``delayed'' refers to neutrinos from muon decay.}
\label{fig:snsisawesome2}
\end{figure}

The SNS produces an intense, isotropic stopped-pion neutrino flux, with a sharply-pulsed timing structure that is highly beneficial for background rejection and precise characterization of the 
backgrounds not associated with the beam~\cite{Bolozdynya:2012xv}.  
See Fig.~\ref{fig:snsisawesome2}.

A neutrino flux simulation has been developed for neutrino experiments at the SNS using GEANT4.10.1~\cite{Agostinelli:2002hh}.  Using this code, a beam of mono-energetic 1-GeV protons impinges upon a liquid mercury target.  
For the baseline simulation, a simplified version of the SNS target geometry was implemented.  
Neutrino spectra and flux are recorded for a COHERENT detector 
in the basement, 20 m from the target center at an angle of $-110^\circ$ with respect to the proton beam direction.
Resulting neutrino spectra and time distributions for different GEANT4 physics lists were compared; good agreement was found 
between ``QGSP\_BERT''  and ``QGSP\_INCLXX'' physics lists.  The flux prediction corresponding to QGSP\_BERT is used for the signal predictions shown here.   See the supplemental materials of Ref.~\cite{Akimov:2017ade} for more details.

The results of the simulations show that the contributions to the neutrino spectrum from decay-in-flight and $\mu$-capture are expected to be very small (Fig. \ref{fig:snsisawesome2}). The contribution to the CEvNS signal from these high-energy neutrinos (E$>$50 MeV) is $<$1\%.  This contamination is at least two orders of magnitude smaller than at other existing facilities, e.g., at Fermilab~\cite{Brice:2013fwa} and J-PARC~\cite{Harada:2013yaa, Axani:2015dha}.  The SNS neutrino flux is also $\sim$80 times larger than that at the Fermilab Booster Neutrino Beam at the same distance from the source.
The overall uncertainty on the neutrino flux is estimated to be 10\%~\cite{Akimov:2017ade}.

\subsection{Background Studies\label{sec:backgroundstudies}}

Understanding and reducing sources of background are critical requirements of the COHERENT experimental program.  The short SNS duty cycle will reduce steady-state backgrounds due to radioactivity and cosmogenics by a factor of $10^3-10^4$. Steady-state backgrounds can also be understood using data taken outside the SNS beam window; however, beam-related backgrounds, especially fast neutrons, will be evaluated using ancillary measurements (described below) and modeling.  

\subsubsection{Neutron Backgrounds}
\begin{figure}[ht]
\centering
\subfigure[]{%
\includegraphics[width=0.45\linewidth]{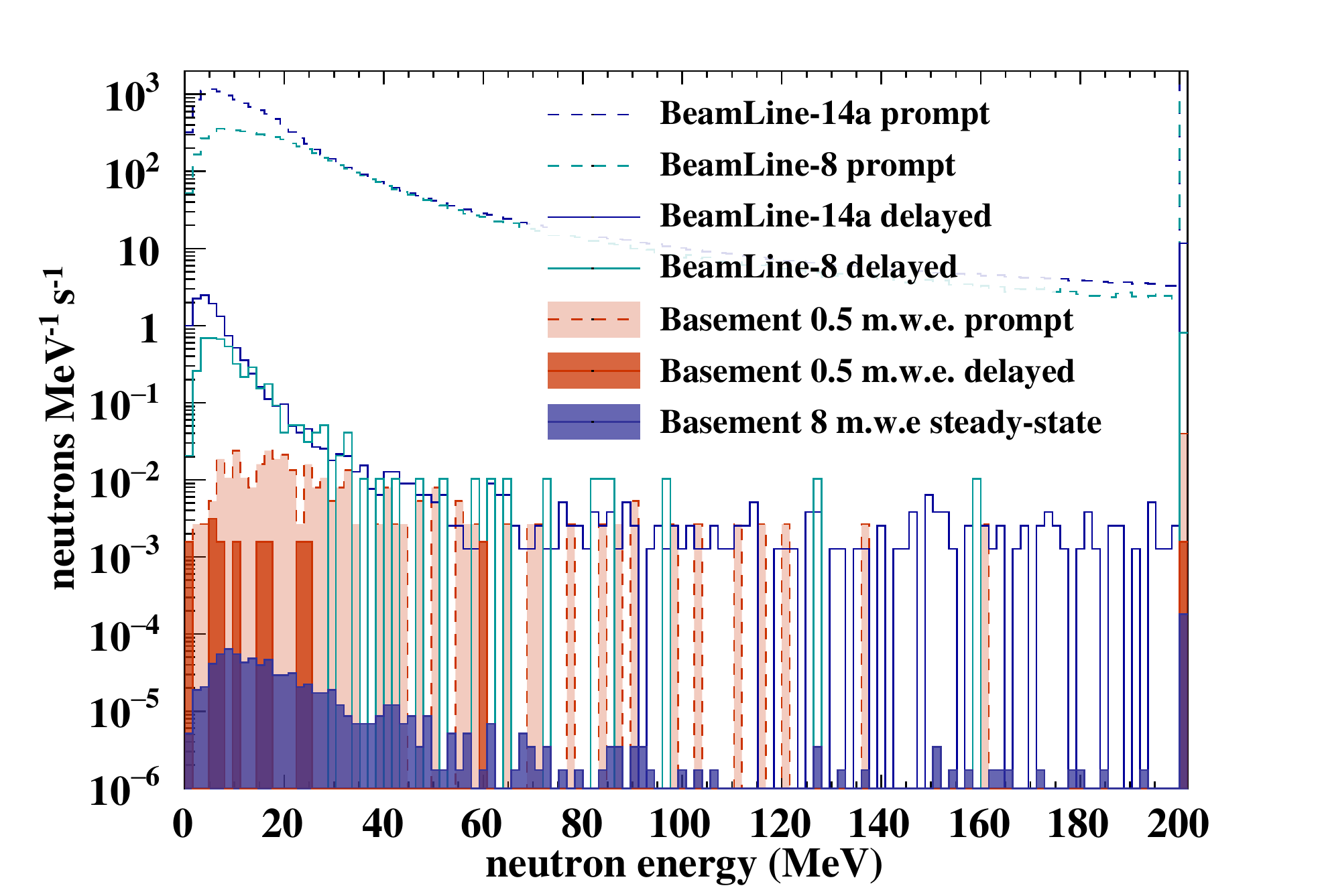}}
\quad
\subfigure[]{%
\includegraphics[width=0.45\linewidth]{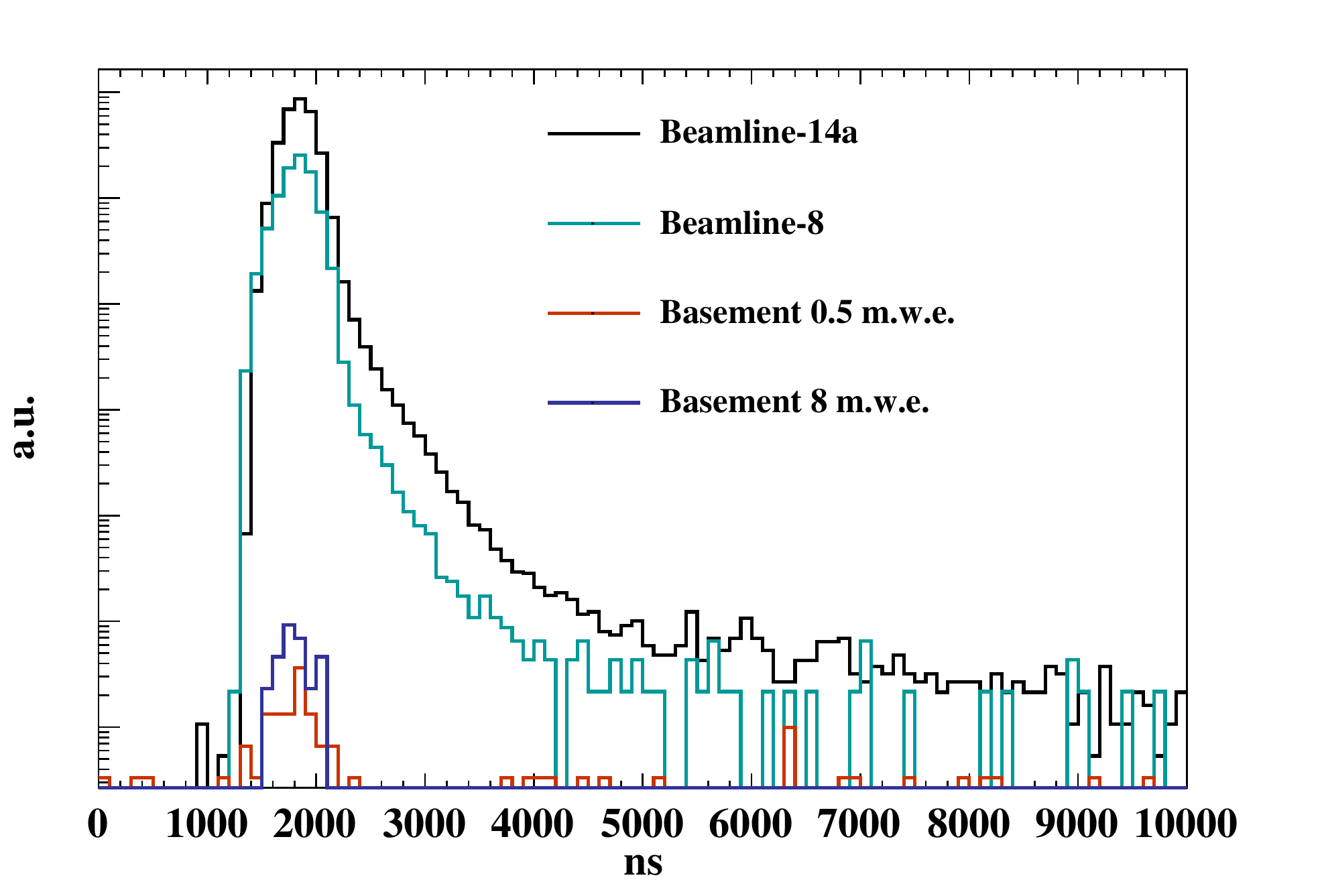}}
\caption{\label{fig:backgrounds_inbeam} (a)~Fast neutron spectra measured with the neutron scatter camera throughout the SNS facility.  A clear reduction by over four orders of magnitude from the experimental hall to the basement locations is seen. No neutron scatters were detected in the delayed window for the basement 8 m.w.e. location. (b) Arrival times of neutrons with respect to SNS beam timing signals.  The proton pulse time offset with respect to zero is 1400~ns. }
\end{figure}

\begin{figure}[ht]
\centering
\subfigure[]{%
\includegraphics[width=0.24\linewidth]{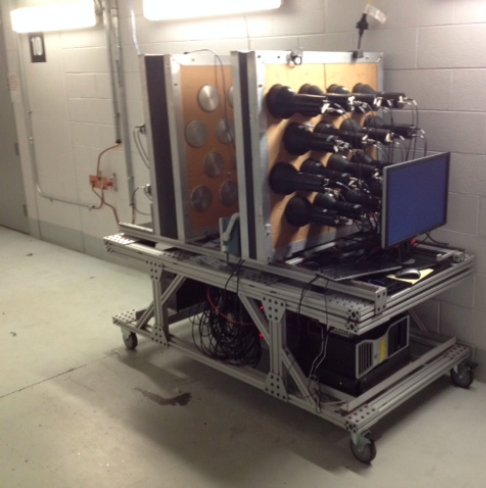}
}
\subfigure[]{%
\includegraphics[width=0.2\linewidth]{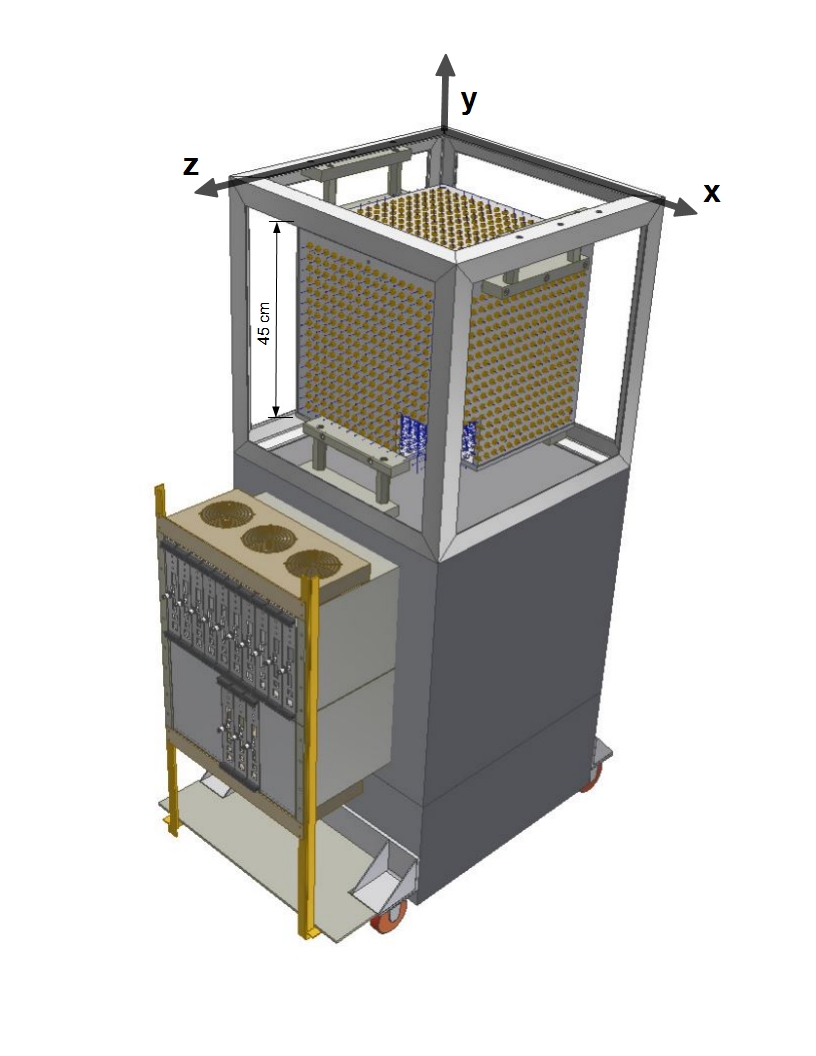}
}
\subfigure[]{%
\includegraphics[width=0.2\linewidth]{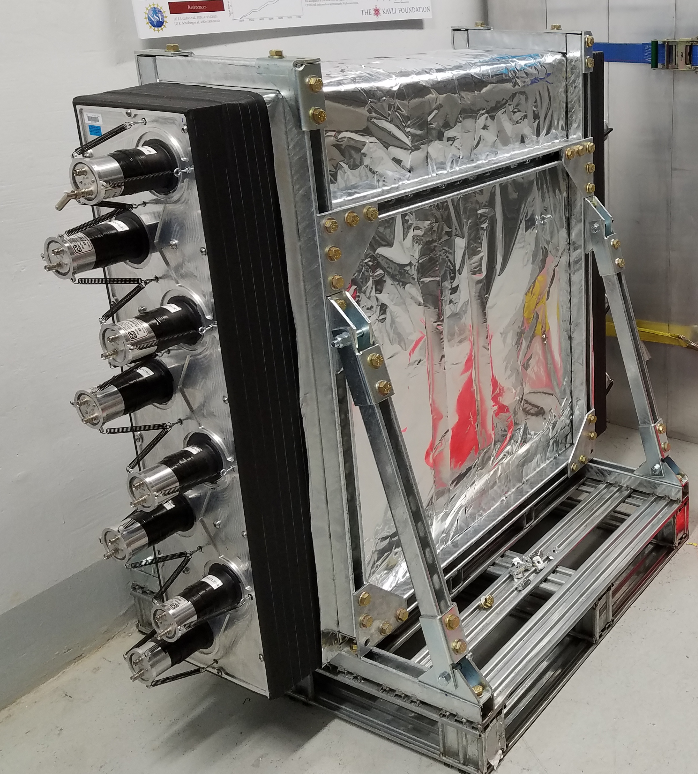}
}
\caption{\label{fig:neutron-detectors} (a) The neutron scatter camera deployed in the SNS basement. (b) The SciBath detector. (c) The MARS detector.}
\end{figure}

A background measurement campaign has been underway since the fall of 2013.  Several detection systems have been used to measure beam-related neutron backgrounds, including: a single portable 5-liter liquid-scintillator detector to assess gross neutron rates at various locations within the SNS target hall and basement, a two-plane neutron scatter camera \cite{Brennan:2009} to provide detailed neutron spectra and some directional information (Fig.~\ref{fig:neutron-detectors}(a)), and a single-plane liquid-scintillator array to provide systematic cross-checks of the neutron scatter camera data.  The measurements taken to date indicate that  Neutrino Alley is very neutron-quiet; the direct beam-related neutrons are more than four orders of magnitude lower in the basement than on the experimental floor of the SNS (see Fig.~\ref{fig:backgrounds_inbeam}). 
The location is also protected from cosmic rays by $\sim$8 meters-water-equivalent (m.w.e.) of overburden.
The measured backgrounds in the basement are used to estimate the beam-related backgrounds for all detector subsystems.

The neutron scatter camera took data at several positions in Neutrino Alley.  In addition, ancillary measurements were performed with the complementary SciBath detector~\cite{Tayloe:2006ct, Cooper:2011kx,Brice:2013fwa} (Fig.~\ref{fig:neutron-detectors}(b)), a liquid-scintillator tracking detector.  The combination of the two provided a systematic cross-check, increasing confidence in the understanding of the neutron background energy spectra.

Beam-related neutron backgrounds can be addressed effectively by taking advantage of the timing characteristics of the SNS (illustrated in Fig. \ref{fig:snsisawesome2}).  The SNS provides beam timing signals that allow precise selection around the $\lsim$800-ns-width arrival time of protons on target. While a small background contribution from beam-related neutrons is expected to add to the $\nu_{\mu}$ signal in the ``prompt'' window (coincident with protons on target), all neutron background measurements indicate that there are negligible beam-related neutrons expected in the ``delayed'' window, when the $\nu_{e}$ and $\bar{\nu}_{\mu}$ from muon decay arrive. 

Acceptance and efficiency corrections to the measured neutron scatter camera spectra were performed in order to provide useful background estimates for the detector subsystem simulations (Fig.~\ref{fig:backgrounds_inbeam}). The flux measurements are dominated by uncertainties due to variations in the angular acceptance of the neutron scatter camera (factors of 3--5).   For the CsI[Na] measurement, neutron backgrounds were evaluated in-situ~\cite{Akimov:2017ade} using a scintillator cell inside the shielding, and in addition were constrained by the absence of inelastic neutron scattering signals from the crystal.

To maintain confidence in the understanding of the neutron backgrounds, continued additional measurements are underway.  The Multiplicity and Recoil Spectrometer (MARS) apparatus~\cite{Roecker:2016juf} (Fig.~\ref{fig:neutron-detectors}(c)), a transportable fast neutron spectrometer using Gd-doped scintillator,  is
currently under commissioning in Neutrino Alley. 

\subsubsection{Neutrino-Induced Neutron Backgrounds}\label{sec:nins}
	The high-energy neutrinos from pion decay at rest have energies above the neutron separation threshold in $^{208}$Pb, a ubiquitous material in detector radiological shields. The  CC interaction ($^{208}$Pb($\nu_e$, e$^-$)$^{208}$Bi), with subsequent prompt neutron emission, may produce significant numbers of \emph{background-producing} neutrons in the Pb shields, pulsed in time with the beam and sharing the 2.2-$\mu$s characteristic time-structure of the $\nu_e$ due to the muon lifetime.  Other isotopes of Pb should have similarly  large neutron-ejection cross sections, and other elements commonly used for shielding (Fe, Cu) may also produce neutrino-induced neutrons (NINs) in CC and NC reactions.
	
The NIN cross section is estimated theoretically only within a factor of $\sim$3~\cite{Fuller:1998kb,Kolbe:2000np,Engel:2002hg}; the only measurement for these targets in this energy range is an inclusive CC $^{56}$Fe($\nu_e$, e$^-$)$^{56}$Co measurement~\cite{Maschuw:1998jf}.  Therefore, a careful measurement of the NIN production cross section is of great importance for background predictions. The measurement of these cross sections also has an impact on the ongoing HALO supernova neutrino detection experiment~\cite{Duba:2008zz, Vaananen:2011bf}. The spallation of neutrons from heavy elements is also expected to influence the nucleosynthesis of heavy elements in supernovae~\cite{qian1997:NINSnucleosynthesis,woosley1990:nuProcess}.

\begin{figure}[ht]
\centering
\subfigure[]{%
\includegraphics[width=0.48\linewidth]{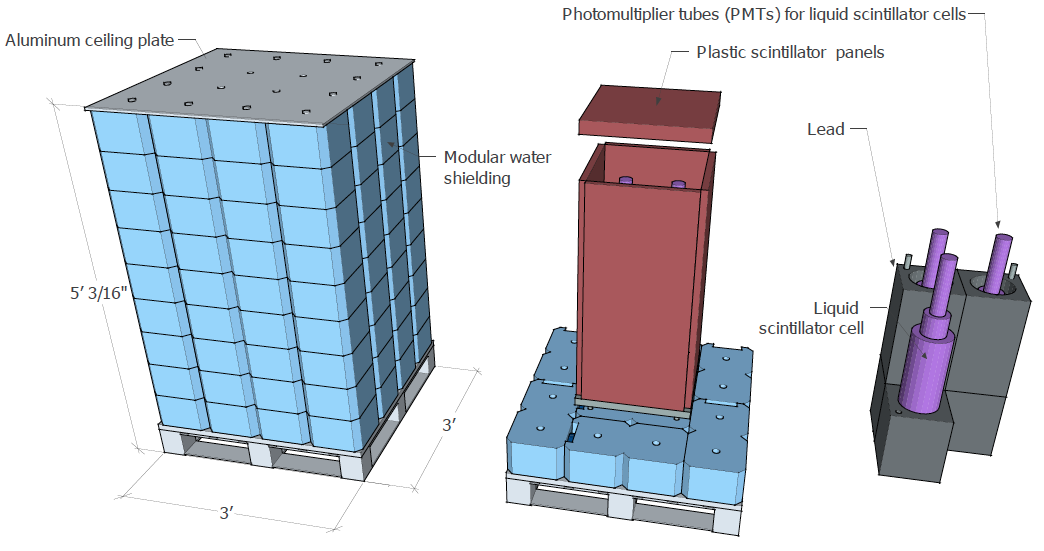}
}
\subfigure[]{%
\includegraphics[width=0.24\linewidth]{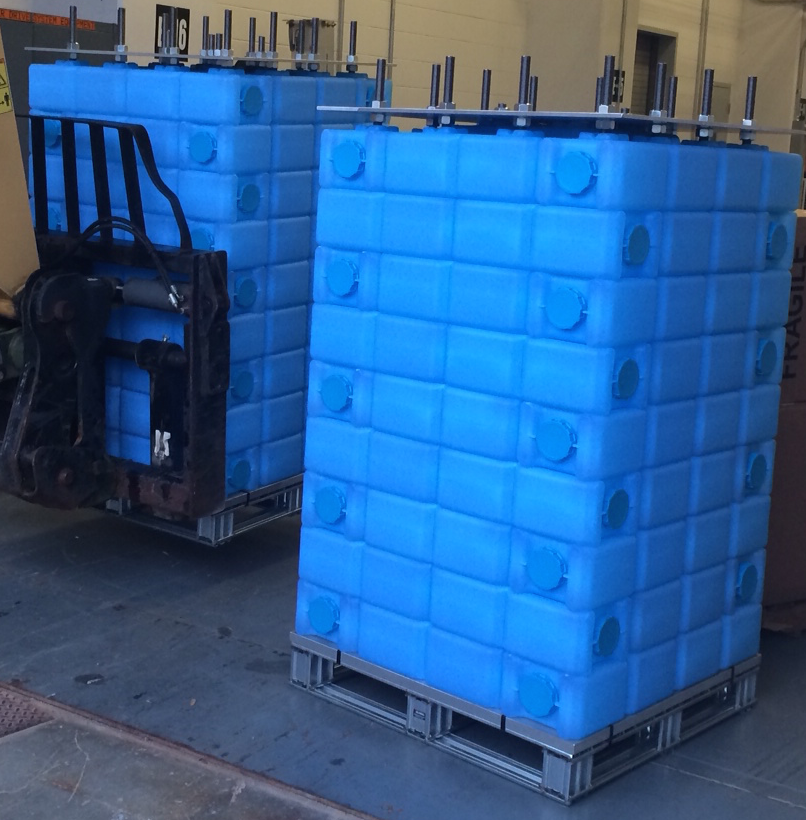}
}
\caption{\label{fig:Nubes} (a) Schematic drawing of the detectors to measure the neutrino-induced neutron cross section on Pb, Fe, and Cu. (b) The imperfectly-named ``Neutrino Cubes,'' or ``NIN Cubes,'' modular neutrino-induced neutron experiments.}
\end{figure}

The COHERENT liquid-scintillator-cell measurement inside the CsI shielding favors a non-zero NIN signal at 2.9$\sigma$~\cite{Akimov:2017ade}, although it also suggests a smaller cross section than from theoretical expectation.
Dedicated apparatuses for NIN measurement containing liquid-scintillator detectors surrounded by lead or other target materials and further surrounded by a muon veto and neutron moderator were designed and deployed to the SNS basement in September 2015 (see Figure~\ref{fig:Nubes}). These detector systems, currently running with lead and iron, are expected to continue operation through the lifetime of the experiment.
COHERENT will use these to measure the production cross sections of NINs in lead, iron, and copper at the SNS, both to evaluate the NIN background for CEvNS and as independent physics measurements.

\subsection{Detector Subsystems} 
The COHERENT Collaboration is deploying four detector subsystems, each containing different target nuclei (summarized in Tab.~\ref{tab:detectors}).  The timing resolution of all four detector subsystems is sufficient to allow the observation of the characteristic 2.2-$\mu$s lifetime of muon-decay neutrinos, a further cross-check that any interactions are due to neutrinos from the SNS. The technologies are mature and all are presently used for direct dark-matter detection or other low-threshold experiments.

\begin{figure}[ht]
\centering
\includegraphics[height=3.5in]{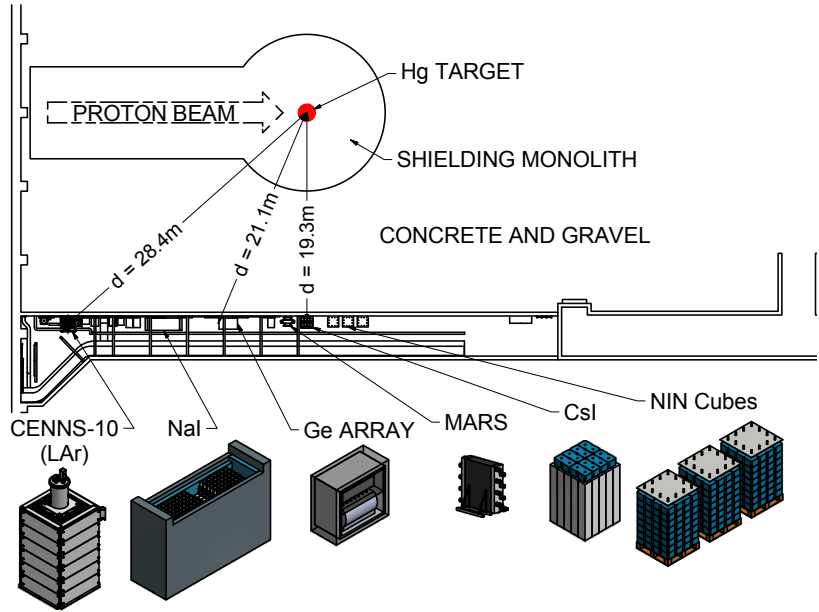}
\caption{Siting of existing and near-future planned detectors in Neutrino Alley. }\label{fig:neutrino_corridor}
\end{figure}

Fig.~\ref{fig:neutrino_corridor} shows the current COHERENT siting in
Neutrino Alley.

\begin{table*}[htbp]
	\centering
 	\begin{tabular}{c|c|c|c|c}
		\hline
 		Nuclear& Technology & Mass & Distance from & Recoil \\

		target & & (kg) &  source (m) & threshold (keVnr)
		\\ \hline 

		CsI[Na] & Scintillating crystal & 14.6 & 19.3 & 6.5 \\ 
		Ge & HPGe PPC & 10 & 22 & 5 \\
		LAr & Single-phase & 22 & 29 & 20 \\
		NaI[Tl] & Scintillating crystal & 185$^*$/2000 & 28 & 13 \\

		\hline
	\end{tabular}
	\caption{\label{tab:detectors}Parameters for the four COHERENT
          detector subsystems. $^*$NaI[Tl] deployed in high-threshold
          mode is described in Sec.~\ref{sec:nai}.}
 \end{table*}

\subsubsection{CsI[Na] Detector Subsystem}

CsI[Na] scintillators (cesium iodide crystals doped with sodium) present several advantages for a CEvNS measurement. This mature technology combines sufficiently low thresholds with large neutron numbers ($N=74, 78$) that make an observation of this process at the SNS feasible. Both recoiling species are essentially indistinguishable due to their very similar mass, greatly simplifying understanding the detector response.  CsI is a rugged, room-temperature detector material, and is also relatively inexpensive ($\sim$ \$1/g).
These detectors have several other practical advantages. CsI[Na] exhibits a high light yield of 64 photons/keVee (electron-equivalent energy deposition) and has the best match to the response curve of bialkali photomultipliers of any scintillator material. CsI[Na] also lacks the excessive afterglow (phosphorescence) that is characteristic of CsI[Tl]~\cite{cosinima}, an important feature in a search involving small scintillation signals in a detector operated at ground level. The quenching factor for nuclear recoils (the fraction of the recoil energy that is detectable as scintillation) in this material over the energy region of interest has been carefully characterized~\cite{cosinima}, using the methods described in~\cite{Collar:2013gu}. The quenching factor has been measured by the Collaboration using quasi-monochromatic neutron beams~\cite{Akimov:2017ade}. 

\begin{figure}[ht]
\centering
\includegraphics[height=2.5in]{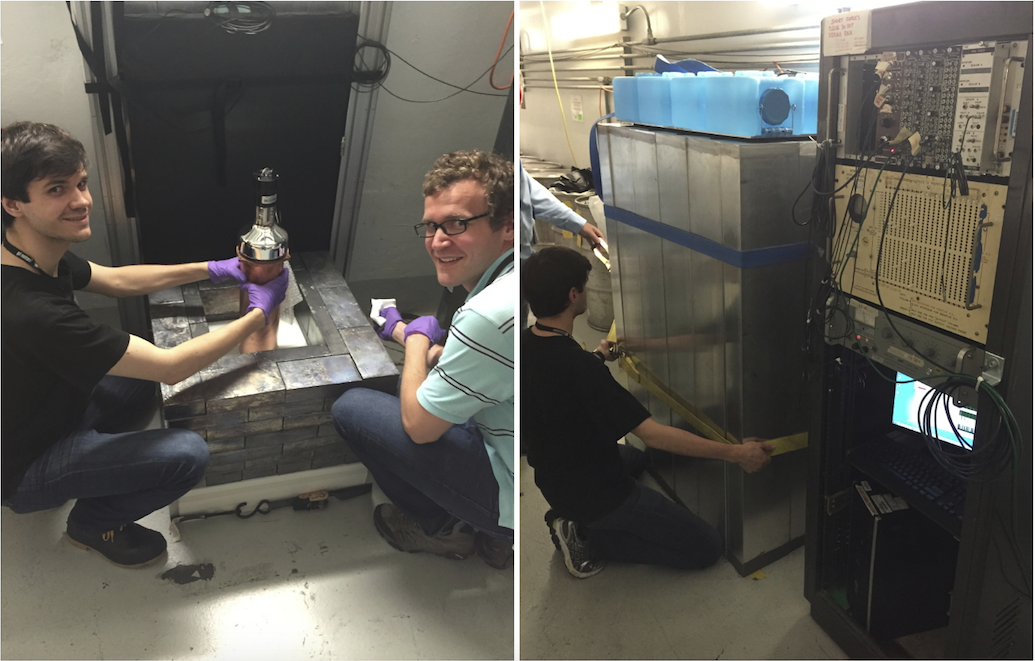}
\caption{The installation of the 14.6-kg, low-background CsI[Na] at the SNS (featuring B.~Scholz~\cite{bjornthesis} and G.~Rich~\cite{graysonthesis}). Successive layers of shielding include (inside-to-out): 7.62 cm HDPE, 15 cm of lead, a 5-cm thick muon veto and 10 cm of water neutron moderator.}\label{fig:csi_installation}
\end{figure}

A 14.57-kg CsI[Na] detector and shielding (Fig.~\ref{fig:csi_installation}) was characterized at the University of Chicago and installed at the SNS in June 2015. 
The shielding consists of  HDPE inner neutron moderator,  Pb (the innermost 5 cm selected for low $^{210}$Pb content), a 99.6\%-efficient muon veto, and an outer neutron moderator and absorber.
See reference~\cite{cosinima} for more details of the detector and its characterization, as well as references~\cite{Akimov:2017ade, graysonthesis, bjornthesis} for details of the deployment and analysis of the 2015-2017 dataset resulting in the first observation of CEvNS.  Continued running will bring improved statistical uncertainty, and analysis improvements are underway to improve signal to background ratio..  The uncertainty on the predicted rate is currently dominated by quenching factor uncertainties, which can be reduced by ancillary measurements.

\subsubsection{P-Type Point Contact HPGe Subsystem}

P-type point contact (PPC) germanium detectors display a set of unique properties in a large-mass
(up to few kg) radiation detector: excellent energy resolution, sensitivity to energy depositions well
below 1~keVee, and the intrinsic radiopurity that characterizes detector-grade (HPGe) germanium
crystals. Following their description in \cite{Barbeau:2007qi,Barbeau:2009zz}, PPCs have been adopted by a number of searches
in dark matter and neutrino physics: CoGeNT (Coherent Germanium Neutrino Technology) \cite{Aalseth:2008, Aalseth:2011, Aalseth:2011wp, Aalseth:2012if}, the \textsc{Majorana Demonstrator}~\cite{Abgrall:2014}, GERDA~\cite{Gerda:2017}, CDEX~\cite{Zhao:2016dak}, TEXONO~\cite{lin2009:texono}, and most recently CONUS~\cite{Lindner:2017,Hakenmuller:2017}.
Continuous development over the course of the past decade has enabled the reproducible manufacture of detectors with masses of up to 2.5~kg, while maintaining an intrinsic electronic noise of $\sim75$~eV~FWHM, equivalent to an energy threshold of $\sim150$~eVee.  
Taking into consideration the measured quenching factor for nuclear recoils in germanium, this allows for detection of recoils with energy below 1~keVnr.
In addition to extremely low energy thresholds, the low noise and 3-eV band-gap energy gives these detectors an excellent energy resolution near threshold. 
As a result, the measured (background-subtracted) energy-deposition spectrum is faithful to the true recoil spectrum, allowing for straightforward searches for deviations of the recoil spectrum due to nuclear form factors or new physics.  

The COHERENT Collaboration aims to deploy a 10-kg array of PPC germanium detectors at the SNS for an initial measurement of CEvNS in germanium and to potentially probe new physics.  COHERENT collaborators have access to 10~kg of existing Canberra BEGe PPCs, which are are currently undergoing refurbishment and being evaluated for internal backgrounds in their original manufacturer-supplied cryostats for potential deployment at the SNS in 2018.  While these on-hand detectors may be capable of an initial CEvNS measurement, they are not representative of the current technological capability in terms of threshold ($\sim$1~keVee as opposed to $\sim150$~eVee) or low-background cryostat fabrication.  An array of four state-of-the-art detectors, each with masses of $\sim2.5$~kg and thresholds comparable to the lowest achieved to date, is being planned for potential deployment at the SNS with the aim of not only sensitively measuring the CEvNS spectrum, but doing so with a recoil energy threshold low enough to probe electromagnetic properties of neutrinos.

\subsubsection{Single-Phase Liquid Argon Subsystem}\label{sec:lar}

The intended first phase of the experiment, proposed in
Ref.~\cite{Akimov:2015nza}, included a two-phase liquid xenon
detector~\cite{Akimov:2012aya}. This detector could not be made
available to COHERENT for this measurement, so an alternative plan
utilizing a single-phase liquid argon detector was developed.
Argon has been proposed for CEvNS measurements at stopped-pion
neutrino sources~\cite{Scholberg:2009ha,Brice:2013fwa} in
the past.  It represents a relatively small value of $N$ and a high
fraction of the zero-spin isotope $^{40}$Ar.  The scintillation light
yield  at 128 nm is high ($\sim$40 photons per keVee). The
different time constants of singlet and triplet molecular decay
channels permit pulse-shape discrimination of nuclear recoils from
electronic recoils~\cite{Boulay:2006}.  
Like other noble liquids, it is naturally clean of radioactive impurities, although it
does contain $\sim$1 Bq/kg of the long-lived cosmogenic isotope $^{39}$Ar which decays via
a 565-keV $\beta$.   However, pulse-shape discrimination together with the pulsed SNS beam provides suppression of steady-state backgrounds.

\begin{figure}[ht]
\centering
\includegraphics[width=0.3\linewidth]{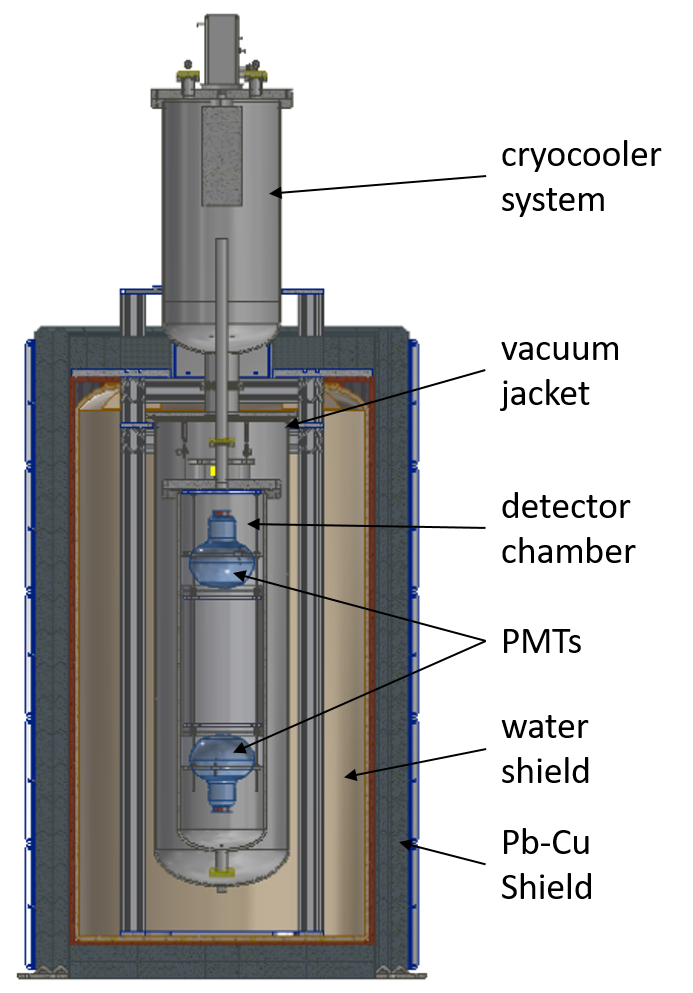}
\includegraphics[width=0.45\linewidth]{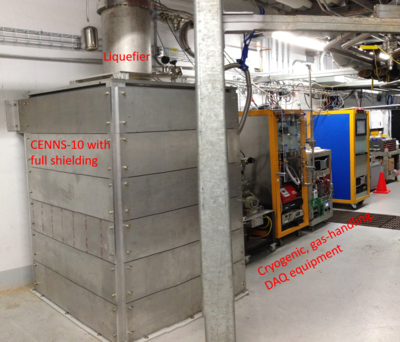}
\caption{Left: Liquid Ar detector section-view schematic.
Right: Liquid Ar detector and associated equipment in Neutrino Alley.}
\label{fig:Ar}
\end{figure}

The currently-installed LAr detector (Fig.~\ref{fig:Ar})~\cite{Tayloe:2017edz} began as the ``CENNS-10'' prototype detector
built at Fermilab~\cite{Brice:2013fwa}, tested and modified at Indiana University, then installed with the requisite shielding in Neutrino Alley at the SNS in December 2016.  It consists of a cylindrical 56.7-liter LAr chamber, viewed by two 8'' Hamamatsu R5912-02MOD photomultiplier tubes
(PMTs) on opposite ends.  The sides of the cylinder are lined by a
tetraphenyl butadiene (TPB)-coated reflector.  This originally
consisted of an acrylic cylinder 27~cm diameter by 46~cm in length
backed by a Teflon sheet. Acrylic disks were placed at the cylinder
ends in front of the PMTs.  In June 2017, the acrylic cylinder
was replaced with a 21-cm diameter Teflon cylinder, the acrylic end
disks were removed and two new PMTs (of same type) with TPB directly coated on the
front surface were installed.   The active fiducial
volume was reduced from 32~kg to 22~kg but it is expected that the light yield improve will compensate. 
The LAr is liquefied and maintained at temperature with a single-stage pulse-tube cold head cryorefrigerator. The boiloff Ar gas is recirculated through a Zr getter to remove N$_2$ to the $\sim 10$-ppm level.  The goal is to obtain several collected photoelectrons per keV for electronic recoils which, with a nuclear recoil response of $\sim 0.25$~\cite{Cao:2014gns}, should allow an energy threshold of $\sim 20$~keV for CEvNS nuclear recoil events. 
 
The detector shielding structure was designed to lower the rate of neutron and photon backgrounds.  It
consists of an inner water tank providing $\sim 22$~cm of water surrounded by 0.6~cm of copper, 10~cm of external lead, and stabilized with 0.6~cm aluminum plates.  Simulations indicate that this structure will reduce the background rate well below that of the 20-100~keV CEvNS nuclear recoil.

\subsubsection{NaI[Tl] Subsystem}\label{sec:nai}

NaI[Tl] is another material with capacity for low-threshold recoil
detection.  The sole stable sodium isotope, $^{23}$Na, with 12 neutrons, has
the lowest $N$ value of COHERENT's target materials, and hence will
result in the highest-energy recoils.  A small-$N$ nuclear target in combination with measurements on heavier nuclides reduces the impact of the flux uncertainty  and improves physics reach (see Sections~\ref{sec:bsm},~\ref{sec:future}). Furthermore, a few-percent
effect from axial contributions is expected at high recoil energy, and
could be of interest to measure in the longer term.  The light yield is
$\sim$40~photoelectrons per keVee.
 
About $\sim$2 tonnes of recycled NaI[Tl] detectors are immediately
available to the COHERENT collaboration, with potentially more (up to
$\sim$9 tonnes) available in the future.
The NaI[Tl] detectors are available in the form of 7.7-kg NaI[Tl] modules
sealed in alumininum and packaged with Burle S83013 (or equivalent)
photomultiplier tubes.  These detectors have rectangular shapes and are
suitable for deployment in a compact array.  Their backgrounds have
been characterized in the range $\sim$ 200 counts keVee$^{-1}$
kg$^{-1}$ day$^{-1}$ in the $\sim$10~keVee recoil energy range.  The higher-energy $^{23}$Na recoils should have better
signal/background than $^{127}$I recoils, so the focus is on these for the
CEvNS measurement.  Quenching factors have been measured to be $\sim$
11\%~\cite{Collar:2013gu}, with recent measurements by COHERENT collaborators at
TUNL favoring a larger $\sim$15\% value. 

The potentially large mass of a detector to be deployed at the SNS
also enables a search for CC
interactions on $^{127}$I, $\nu_e + {}^{127}{\rm I}\rightarrow e^- +
{}^{127}{\rm Xe^{(*)}}$,
which has a lower cross section than CEvNS on CsI
by two orders of magnitude~\cite{Formaggio:2013kya,Engel:1994sq}, but which produces high-energy
electrons (MeV scale) which are easily observable.
The $\nu_e$ cross section for interaction on $^{127}$I,  previously proposed for solar neutrino studies~\cite{Haxton:1988bk}, has been measured previously with 34\% uncertainty at a stopped-pion
source~\cite{Distel:2002ch}.  Improved cross-section measurements of neutrino-induced interactions with significant momentum transfer could provide a new handle on possible $g_A$ quenching, a matter of critical importance for future neutrinoless double-beta decay experiments (e.g.,~\cite{Engel:2016xgb}). In particular, recent nuclear models predict a dependence of $g_A$ quenching on momentum transfer (e.g.,~\cite{Menendez:2011qq}) that could be tested with a stopped-pion neutrino source.  We hope to make an improved measurement of the $^{127}$I CC cross section and to explore the sensitivity of the NaI array to $g_A$-quenching physics. Future experiments using double-beta decay isotopes as targets could also provide interesting constraints on nuclear matrix element calculations~\cite{Volpe:2005iy}.

\begin{figure}[ht]
\centering
\includegraphics[width=0.35\linewidth]{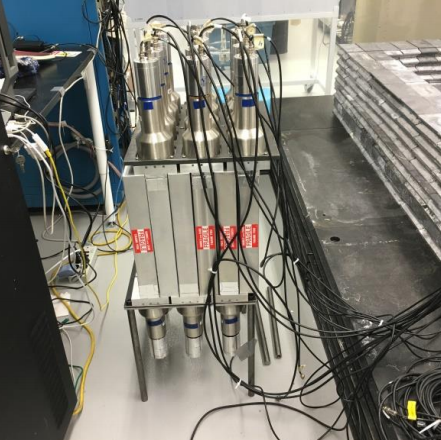}
\includegraphics[width=0.28\linewidth]{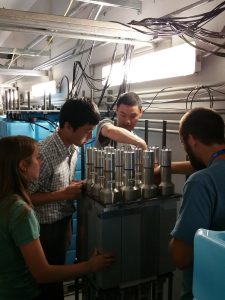}
\caption{Left: NaI[Tl] test stand at Duke University, showing individual detectors.
Right: installation of the 185-kg NaI array in Neutrino Alley.}
\label{fig:NaI}
\end{figure}

The photomultiplier tubes packaged with the existing NaI[Tl] detectors
have insufficient gain to observe a CEvNS signal with high efficiency.
Hence, the Collaboration is refurbishing the PMT
bases to allow higher-gain running, in order to 
enable observation of lower-energy recoils. 
A total of 185~kg of NaI[Tl] detectors have been deployed at the SNS
(see Fig.~\ref{fig:NaI})
and have been running since November 2016 to measure backgrounds and explore CC interactions.
Several upgrades, including improved shielding and cosmic-ray veto
panels, have recently been completed.  In the next phase, 2 tonnes will be deployed in
high- or dual-gain mode with potential sensitivity to CEvNS signals as well as CC interactions.  Although backgrounds are expected to be higher than for the other detectors, the large amount of target mass will provide large statistics.

\section{COHERENT Status and Upgrades}\label{sec:status}

\begin{figure}[ht]
\centering
\includegraphics[height=2.5in]{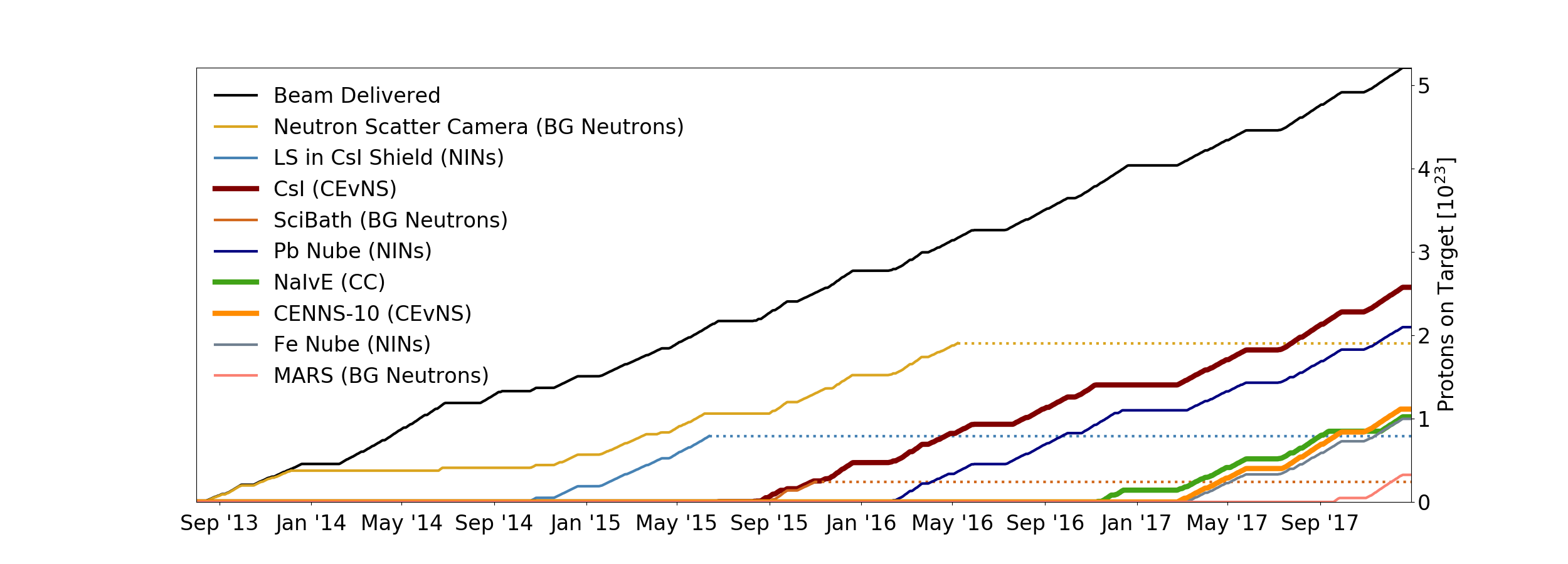}
\caption{Integrated protons on target from the SNS delivered to different COHERENT subsystems.   The Neutron Scatter Camera and SciBath detectors are described in Sec.~\ref{sec:backgroundstudies}.  The ``LS in CsI Shield'' line refers to the liquid scintillator cell deployed in the CsI detector shield described in Ref.~\cite{Akimov:2017ade}.  The Pb and Fe Nubes are described in Sec.~\ref{sec:nins}. ``NaIvE (CC)'' refers to the NaI[Tl] 185-kg detector running in low-gain CC mode as described in Sec.~\ref{sec:nai}. ``CENNS-10'' refers to the single-phase liquid argon detector described in Sec.~\ref{sec:lar}.  ``MARS'' refers to the neutron detector described in Sec.~\ref{sec:backgroundstudies}.}\label{fig:pot}
\end{figure}

As of this writing, the SNS-delivered protons on target for different COHERENT subsystems (corrected for detector livetime)  are shown in Fig.~\ref{fig:pot}.

The status of the four CEvNS detector subsystems at the time of this writing are as follows:
\begin{description} 
  \item[] \textbf{CsI[Na]:} The detector was installed in 2015 and provided data for the first result in 2017~\cite{Akimov:2017ade}.  It will continue to run in the near future.  Statistical uncertainties will be reduced, and we expect analysis improvements as well as improvement of quenching factor uncertainty from ancillary measurements.

\item[] \textbf{LAr:}  The single-phase LAr scintillation detector was installed in Fall 2016 and took first beam in December 2016, running in that configuration through May 2017.  It was upgraded for improved light yield for the next run period beginning July 2017.  It will continue to run in that configuration into 2018. 
  
\item[]\textbf{NaI[Tl]:} An initial deployment of 185 kg of NaI began data-taking in late 2016 in high-threshold mode (sensitive to CC on $^{127}$I).  Minor upgrades are ongoing, and refurbishment of PMT bases is underway to enable CEvNS running for additional detector deployment.
 
  \item[] \textbf{Ge:}  The available Ge detectors are under test and refurbishment, with installation expected in 2018.  The Collaboration is seeking to purchase and deploy additional detectors.

\end{description}

\subsection{Future Upgrades}\label{sec:upgrades}

The COHERENT Collaboration is pursuing multiple possibilities for
future upgrades to the suite of detectors in Neutrino Alley.
Additional detector mass is under consideration for all existing targets except CsI[Na].

\begin{itemize}
\item A tonne-scale LAr detector is under consideration.
\item Additional Ge detectors may be added to the existing apparatus.
\item NaI[Tl] detectors, up to 9 tonnes, if available, may be added.
\end{itemize}

In addition, detectors employing different targets are also under
consideration.  There is benefit for physics sensitivity in measuring
CEvNS with $N$ values as different from each other as possible; a case
in point is Ne and Xe, with greatly improved sensitivity to
NSI~\cite{Scholberg:2005qs}.  Other nuclei are of interest for neutron-radius measurements~\cite{Cadeddu:2017etk}.

Also under consideration by the Collaboration is ``Mini-HALO'' for a
dedicated measurement of the NIN-production cross section and detection efficiency
for the HALO experiment~\cite{Duba:2008zz}.

The dominant systematic uncertainty for many targets is due to the
quenching factor uncertainty, and ancillary measurements are underway for
independent measurements of these in the current COHERENT targets and others.

Although not dominant for the current CsI CEvNS measurement,
uncertainties on the neutrino flux (currently $\sim$10\%) will
eventually become dominant.  A current plan is to address this with
heavy water, because the theoretical uncertainty for neutrino interactions
on deuterons is small ($\sim 2-3\%$)~\cite{Kozlov:1999ct,
  Nakamura:2002jg,Formaggio:2013kya}.  A tonne-scale D$_2$O
flux-monitoring detector is under design.  Such a detector filled with regular water can also measure CC and NC interactions on $^{16}$O, of relevance to measurements of supernova neutrinos in Super-Kamiokande and Hyper-Kamiokande~\cite{Scholberg:2012id}.

\section{Summary}

In summary, the COHERENT program will take advantage of the extremely
high-quality stopped-pion neutrino source available at the Spallation
Neutron Source at Oak Ridge National Laboratory for new CEvNS
measurements. Neutrino Alley in the basement of the SNS target
building provides a unique low-background environment for neutrino
experiments.  Following the first measurement of CEvNS in CsI, four
detector subsystems, based on CsI, Ge, Ar, and NaI, will demonstrate
the $N^2$ dependence of the cross section and significantly improve
constraints on non-standard interactions of neutrinos with nuclei.  A
secondary goal of COHERENT is to measure the cross sections for NINs
on lead, iron, and potentially other targets, using independent
scintillator detectors, in order to estimate the NIN component of the
CEvNS detector signal.  These measurements will be of interest in
their own right to the nuclear-physics and particle-astrophysics
communities, due to their relevance for supernova-neutrino detection.
Beyond the current program, there is physics motivation for several
potential upgrades, including increased mass for any of the existing
targets, additional targets, and flux monitoring.

\vspace{0.1in}
\noindent
\textbf{Acknowledgments}

We acknowledge support from: the Alfred P. Sloan Foundation (BR2014-037), the Consortium for Nonproliferation Enabling Capabilities (DE-NA0002576), the Institute for Basic Science (Korea) (IBS-R017-G1-2017-a00), the National Science Foundation (PHY-1306942, PHY-1506357, PHY-1614545, HRD-1601174), Lawrence Berkeley National Laboratory Directed Research and Development funds, Russian Foundation for Basic Research (No. 17-02-01077\_a), the Russian Science Foundation in the framework of MEPhI Academic Excellence Project (contract 02.a03.21.0005, 27.08.2013), Sandia National Laboratories Directed Research and Development Exploratory Express Funds, Triangle Universities Nuclear Laboratory, the U.S. Department of Energy Office of Science (DE-SC0009824, DE-SC0010007, Early Career Award DE-SC0014249, DE-SC0014558),  Office of Science Graduate Student Research (SCGSR) program,
the U.S. Department of Energy’s National Nuclear Security Administration Office of Defense, Nuclear Nonproliferation Research, and Development, and the University of Washington Royalty Research Fund (FA124183).
Sandia National Laboratories is a multimission laboratory managed and operated by National Technology and Engineering Solutions of Sandia LLC, a wholly owned subsidiary of Honeywell International Inc. for the U.S. Department of Energy's National Nuclear Security Administration under contract DE-NA0003525. 
This work was supported in part by the Kavli Institute for Cosmological Physics at the University of Chicago through grant NSF PHY-1125897, and an endowment from the Kavli Foundation and its founder Fred Kavli.
This work was sponsored by the Laboratory Directed Research and Development Program of Oak Ridge National Laboratory, managed by UT-Battelle, LLC, for the U. S. Department of Energy. It used resources of the Spallation Neutron Source, which is a DOE Office of Science User Facility. This material is based upon work supported by the U.S. Department of Energy, Office of Science, Office of High Energy Physics.
This research uses resources of the Oak Ridge Leadership Computing Facility, which is a DOE Office of Science User Facility. Raw experimental data are archived at its High Performance Storage System, which provides 263 TB of COHERENT dedicated storage. We are
grateful for additional resources provided by the research computing centers at the University of Chicago, and Duke University.  We thank Pacific Northwest National Laboratory colleagues for CsI[Na] detector
contributions.  We thank Fermilab for the loan of the CENNS-10 detector.

\bibliographystyle{vitae}
\bibliography{refs}

\end{document}